\documentclass[11pt]{article}
\usepackage{jheppub}
\usepackage{graphicx}
\usepackage{enumerate}
\usepackage[titletoc, title]{appendix}

\graphicspath{{./Artwork/}}

\newtheorem{theorem}{Theorem}
\newtheorem{suspicion}{Hypothesis}

\title{On powercounting in perturbative quantum gravity theories through color-kinematic duality}
\author{Rutger H. Boels}\emailAdd{Rutger.Boels@desy.de}
\author{and Reinke Sven Isermann}\emailAdd{Reinke.Sven.Isermann@desy.de}
\affiliation{II. Institut f\"ur Theoretische Physik, Universit\"at Hamburg\\ Luruper Chaussee 149, D- 22761 Hamburg, Germany }


\abstract{The standard argument why gravity is not renormalisable relies on direct powercounting of Feynman graphs to estimate the degree of UV divergence. In several (highly) supersymmetric examples the actual divergences have been shown to be considerably better. In these examples the improvement follows from a conjectured duality between color and kinematics. In this paper we initiate the systematic study of quite general powercounting under the assumption that color-kinematic duality exists. The main technical tool is a reformulation of the duality in terms of linear maps, modulo subtleties at loop level mostly inherent to the duality. This tool may have wider applications in both gauge and gravity theories, up to resolution of the subtleties. Here it is first applied to the large Britto-Cachazo-Feng-Witten (BCFW) shift behavior of gravity integrands constructed through the duality. Assuming color-kinematic duality and reasonable technical requirements hold these shifts are shown to be independent of loop order, which is a new argument for massive cancellations with respect to the Feynman graph expression. More speculatively, the same approach is then applied to provide estimates of the overall degree of UV divergence in quite general gravity theories, assuming the duality exists. The developed arguments apply to all multiplicity. The manifest cancellations obtained in these estimates depends on the exact implementation of the duality at loop level, especially on graph topology. Finally, some evidence for the duality to all loop orders is provided from an analysis of BCFW shifts of gauge theory integrands through Feynman graphs. }

\begin{document}

\maketitle

\section{Introduction}
Perhaps the largest unsolved problem in theoretical physics is the unification of the theory of general relativity with the principles of quantum mechanics. A range of possible solutions to this problem have been proposed. These typically involve either embedding into a much larger theory (e.g. string theory), appealing to non-perturbative dynamics (asymptotic safety) or by starting with a radically different formulation of the theory (loop quantum gravity), see e.g. \cite{Woodard:2009ns} for a review. None of these have to date led to a phenomenologically viable and/or theoretically well-understood theory of quantum gravity.

It is well known that classically Einstein's gravity theory can be formulated as a Lagrangian field theory, just as the standard model of particle physics can be. However, the latter can be turned into a well-defined quantum field theory by perturbative quantization. Divergent integrals which arise in this process in the standard model can be given a physical meaning through renormalization: all divergences can systematically be absorbed into redefinitions of coupling constants and field normalizations. For renormalization to work the divergences must be under control and well understood. In particular for predictive power renormalization should involve only a \emph{finite} number of coupling constants: this implies a finite number of measurements can fix these constants and all further measurements are predictions.

The usual argument why gravity cannot be quantized according to the rules of quantum field theory involves a na\"ive counting of powers of loop momenta. Around a flat background metric a perturbation theory in the coupling constant can be formulated just as it can in a gauge theory leading to Feynman graphs. When calculating say an $n$-point scattering amplitude at $l$ loops one encounters complicated divergent integrals, roughly of the type 
\begin{equation}
\sim \int d^D{l} \frac{l^{\mu_1} \ldots l^{\mu_{k}}}{l_1^2 \ldots l_{i}^2 }
\end{equation}
where $i$ is the number of massless propagators in this particular loop. The maximal number of loop momenta in the numerator, corresponding to the most divergent integrals in gravity theory, is $k=2 i$. In contrast, in Yang-Mills theories this number is $k=i$: the integrals appearing in gravity theories are therefore much more divergent in the ultraviolet "$l_i\rightarrow \infty$" limit than their gauge theory equivalents. This difference can simply be traced to the fact that in four dimensional gauge theory the coupling constant is dimensionless, while it has a negative mass dimension in gravity in this number of dimensions. In particular this implies that in a typical calculation in gravity there are ever more divergent integrals appearing at each consecutive loop order, na\"ively requiring an infinite amount of coupling constant and field renormalizations. If this number is truly infinite a perturbative quantum theory has no predictive power. In other words, there is no sensible implementation of renormalization known in gravity theories. There are two related ways out of  this which preserve renormalizability: the contributions of the divergent integrals could sum to zero or the divergences are not independent. In the latter scenario a symmetry would relate the divergent coefficients\footnote{An example of this are supersymmetric gauge theories on standard $\mathcal{N}=1$ superspace: the field of mass dimension zero appearing in the vector multiplet can always be gauge transformed away provided the full supersymmetric gauge symmetry is present in the Lagrangian. }.

Investigating either scenario just mentioned is technically difficult in a Feynman graph approach in any gravity theory: there are many graphs, each with complicated expressions even for the simplest tree level calculations. At loop level this leads to complicated sums over divergent integrals. Using textbook methods it is therefore prohibitively difficult to find any would-be pattern within the divergences of gravity theories.  Despite this, it has been verified non-zero divergences arise at one loop in generic matter-coupled gravity \cite{'tHooft:1974bx} and at two loops in pure Einstein gravity \cite{Goroff:1985th}.  In supersymmetric theories of gravity no explicit divergence has ever been demonstrated\footnote{Recent results of explicit calculations include four point amplitudes in $\mathcal{N}=8$ up to four loops \cite{Bern:2009kd}, five point amplitudes up to two loops \cite{Carrasco:2011mn}. Beyond this there are firm arguments the first divergence appears at seven loops, see \cite{Beisert:2010jx}  and also \cite{Vanhove:2010nf}. In $\mathcal{N}=4$ explicit calculation has reached three loops, four points. See \cite{Bern:2012cd}, \cite{Bern:2012gh}, see also \cite{Tourkine:2012ip} for a more general argument.}. Beyond explicit finite order results powercounting in these theories remains therefore a rough upper bound on the apparent degree of ultraviolet divergence of gravitational theories: cancellations could very well be hidden within the sums. Until very recently however no mechanism was known to drive and study these would-be cancellations in loop diagrams. 

This has changed with the work of Bern, Carrasco and Johannson (BCJ)  \cite{Bern:2008qj}, \cite{Bern:2010ue}. These authors have proposed that at any loop order the integrand of gravity theories can in a precise sense be constructed as a double copy of two gauge theories. The field content of the gravity theory is simply the tensor product of the gauge theory factors. The driving force is a certain duality between color and kinematics which gives the kinematical factors the structure of a Lie algebra. The primary motivation for this type of construction at tree level are the Kawai-Lewellen-Tye (KLT) \cite{Kawai:1985xq} relations in string theory where they are basically an observation about the relation between open and closed string correlation functions. The status of color-kinematic duality is reviewed in section \ref{sec:rev}. 

Since gauge theory amplitudes are well-behaved in the ultraviolet, a question arises if the BCJ representation of the gravity integrand as a double copy of gauge theory implies any generic cancellations with respect to na\"ive powercounting. This question has been studied in specific example amplitudes, see footnote 2, up to relatively high (four) loop order. Although this progress is impressive it is also clear that the loop-by-loop approach of these examples will never lead to a proof of finiteness of any theory as this requires all order results. In this article a systematic study is initiated of the cancellations color-kinematic duality implies within quite general gravity theories to all loop orders in the gravity integrand. To this end we reformulate color-kinematic duality into a problem involving linear maps and study limits of this problem. The assumption that color-kinematic duality holds translates into the assumption a solution to this problem exists. The linear algebra approach initiated here could be of wider interest: in this paper it is used to study two examples of powercounting to all loop orders. The first is the study of large so-called Britto-Cachazo-Feng-Witten (BCFW) shifts of the gravity integrand, while the second makes some inroads into powercounting the overall degree of ultraviolet divergence directly.

This article is structured as follows. First some of the necessary background material will be reviewed in section \ref{sec:rev}. In section \ref{sec:amplitudes} it will be shown that the improved scaling of tree level gravity amplitudes follows directly from color-kinematic duality. In section \ref{sec:integrands} it will then be argued that if color-kinematic duality holds at any loop order in addition to some technical assumptions, then the gravity integrand scales the same as the tree level amplitude. More speculatively, the same techniques will be applied to study powercounting aimed directly at UV divergences in section \ref{sec:uv}.  In section \ref{sec:feynman} it will be shown how this scaling behavior can to some extent also be understood from Feynman graphs directly. The discussion and conclusion section \ref{sec:discussion} follows. Some technical details are contained in the appendices.


\section{Review of concepts}\label{sec:rev}
This section contains a brief overview over several concepts needed in the main body of this work.

\subsection{Color-kinematic duality}
Tree level closed string amplitudes in a flat background can be related to a certain sum over products of open string amplitudes. This is the content of the KLT relations \cite{Kawai:1985xq}, which basically follow directly from the CFT picture of the string theory and the factorization of closed string vertex operators into left and right sectors. In the field theory limit these relations relate gravity amplitudes to a ``square'' of gauge theory amplitudes: a sum over products of gauge theory amplitudes with momentum-dependent constants. From a Lagrangian point of view these relations have always been quite mysterious. Recently, the relations have been recast in a more streamlined form by Bern, Carrasco and Johansson (BCJ)  \cite{Bern:2008qj} at tree level and conjecturally extended to the loop level integrand \cite{Bern:2010ue}. In this form the relations are a consequence of a certain duality between color and kinematics. 

To expose color-kinematic duality amplitudes in a general gauge theory coupled to adjoint matter in $D$ dimensions areÄ rewritten as an expression in terms of cubic graphs only.  Let $\Gamma_i$ denote the set of all possible connected cubic graphs which can be drawn for a scattering amplitude with $n$ external legs. Then for each graph in this set one constructs an associated color factor $c_i$ by combining the structure constants of the three vertices in the obvious way. In this way the $n$-point Yang-Mills tree level amplitude can be written as, 
\begin{equation}\label{eq:BCJtree}
\mathcal{A}_n= g_{ym}^{n-2} \sum_{\Gamma_i}\frac{n_i c_i}{s_i}
\end{equation}
Here the coefficients $n_i$ are referred to as kinematic numerators. $s_i$ is the product of all propagators naturally associated to each cubic graph. This amounts to a rewriting of gauge theory amplitudes, as for instance in any gauge one could reabsorb quartic vertices into effective cubic ones according to their structure constant structure. 

\begin{figure}[t]\label{fig:IHX}
\centering
\includegraphics[scale=0.6]{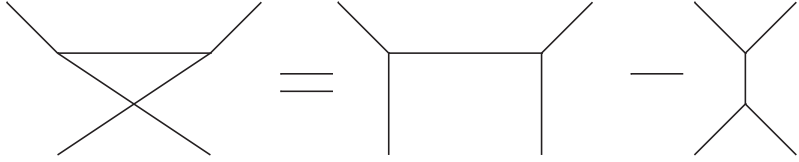}
\caption{A graphical illustration of the Jacobi relations.}
\end{figure}

The first non-trivial step in the Bern-Carrasco-Johansson approach is to demand that whenever color factors obey a Jacobi relation, the corresponding kinematic numerators must do so as well, i.e. 
\begin{equation}\label{eq:colorkinematicduality}
c_i=c_k-c_j \Rightarrow {n}_i={n}_k-{n}_j
\end{equation}
It is useful to note that this identity can be given a nice graphical form depicted in figure \ref{fig:IHX}. 
To state this first part of color-kinematic duality explicitly:
\begin{theorem}\label{theo:BCJpart1tree}
For every tree level amplitude in a theory with adjoint matter only one can find numerators $n_j$ such that they satisfy the Jacobi relations of equation \eqref{eq:colorkinematicduality} and reproduce the amplitudes through \eqref{eq:BCJtree}.
\end{theorem}
These numerators will be termed ``color-dual''. The representation of a gauge theory amplitude using color-dual numerators in equation \eqref{eq:BCJtree} will be called BCJ representation in the following.

The numerators in equation \eqref{eq:BCJtree} are not unique: for every set of shifts $\Delta_i$ such that
\begin{equation}
\sum_{\Gamma_i}\frac{\Delta_i c_i}{s_i} = 0\label{gaugecondition}
\end{equation}
the numerators given by
\begin{equation}
n'_i = n_i+\Delta_i
\end{equation} describe the same amplitude through equation \eqref{eq:BCJtree}. If the shifts $\Delta_i$ also obey equation \eqref{eq:colorkinematicduality}  the new numerators satisfy the Jacobi relations if the old ones did. This freedom in shifting numerators will be referred to as generalized gauge transformations. To some extent this shifting freedom is a manifestation of the fact that the color factors $c_i$ are reducible: they obey (many) Jacobi relations. It will be shown below there is a bit more to it than this. The usual gauge freedom of amplitudes is a subset of the generalized gauge transformation. 

A generic set of numerators $n'_i$ which describe the amplitude through equation \eqref{eq:BCJtree} will not be color-dual, i.e. 
\begin{equation}\label{eq:moregengaugetrafo1}
c_i=c_k-c_j \Rightarrow {n}'_i-{n}'_k+{n}'_j = u_m
\end{equation}
will hold. However, in theories for which there are numerators $n_i$ which obey \eqref{eq:colorkinematicduality} there will be a generalized gauge transformation $\Delta_i$ for which
\begin{equation}\label{eq:moregengaugetrafo2}
c_i=c_k-c_j \Rightarrow \Delta_i-\Delta_k+\Delta_j = u_m \qquad \sum_{\Gamma_i}\frac{\Delta_i c_i}{s_i} = 0
\end{equation}
holds. In this article when not explicitly stated otherwise generalized gauge transformations will be color-dual. 

The second step in color-kinematic duality is the following
\begin{theorem}\label{theo:BCJpart2tree}
Given a set of kinematic numerators which satisfy the kinematic Jacobi relations one can construct a gravity amplitude $M$ by replacing the color factor in equation \eqref{eq:BCJtree} by another set of kinematic numerators
\begin{equation}
\mathcal{A}_n= g_{ym}^{n-2}\sum_{\Gamma_i}\frac{{n}_i c_i}{s_i} \Rightarrow M_n = \kappa^{n-2} \sum_{\Gamma_i}\frac{{n}_i \tilde{{n}}_i}{s_i}
\label{eq:squaringrelation}
\end{equation}
with $\kappa$ the gravitational coupling constant. The field content of the gravitational theory is the tensor product of the field content of the two gauge theory copies. 
\end{theorem}

The generalized gauge invariance guarantees that shifts $\Delta_i$ of the kinematic numerators which satisfy \eqref{gaugecondition} leave the squaring relations in equation \eqref{eq:squaringrelation} invariant. The reason is that the only input in \eqref{gaugecondition} for the color factors is the algebraic Jacobi identity and the numerators are assumed to satisfy this identity. Note that the gauge invariance allows one to have only one set of numerators in equation \eqref{eq:squaringrelation} be explicitly color-dual. The other can be related to an explicitly color-dual set by a generalized gauge transformation as in equations \eqref{eq:moregengaugetrafo1} and \eqref{eq:moregengaugetrafo2}. 

The field content of the two copies of gauge theory does not have to be the same. For instance with both gauge theory numerators taken from $\mathcal{N}=4$ the BCJ construction gives $\mathcal{N}=8$ supergravity, while with numerators in pure Yang-Mills the resulting theory is that of a graviton, dilaton and a two-form. The latter theory is sometimes  referred to as $\mathcal{N}=0$ supergravity. Taking one set of numerators from pure Yang-Mills and the other from $\mathcal{N}=4$ SYM gives $\mathcal{N}=4$ SUGRA. A more complete classification for $D=4$ can be found in \cite{Damgaard:2012fb}. 

The two steps above are theorems since explicit sets of color-dual numerators have been found for instance in \cite{BjerrumBohr:2010hn,  Kierm:2010xq, Mafra:2011kj}. Furthermore, under the assumption that there are no non-trivial poles hiding in the set the double copy of equation \eqref{eq:squaringrelation} has been proven in \cite{Bern:2010yg}.  A construction for numerators involving self-dual Yang-Mills theory has appeared in \cite{Monteiro:2011pc} as well as a construction in terms of auxiliary scalar field theories in \cite{BjerrumBohr:2012mg} which will be discussed below.  It has been shown at tree level for self-dual Yang-Mills theory that the symmetry structure implied by color-kinematic duality can be understood in terms of a hidden infinite kinematic Lie algebra \cite{Monteiro:2011pc, BjerrumBohr:2012mg} which basically amounts to diffeomorphism invariance in a restricted part of space-time. Recently, there has also been work on a version of color-kinematic duality in three dimensions \cite{Bargheer:2012gv, Huang:2012wr}. Moreover, there has also been work on the color-kinematic duality involving polygons of MHV amplitudes \cite{Yuan:2012rg}.

As a conjecture color-kinematic duality has been extended to loops at the level of the integrand \cite{Bern:2010ue}. The starting point is as at tree level to express the integrand of a gauge theory amplitude at a fixed loop order $l$ as a sum over trivalent graphs,
\begin{equation}\label{eq:BCJintegrand}
 \mathcal{A}^l_n= g_{ym}^{n-2+2l}  \int \prod_{j=1}^ld^DL_j \sum_{\Gamma_i} \frac{1}{S_i} \frac{n_i c_i}{s_i}
\end{equation}
Here $D$ is the space-time dimension and $S_i$ indicates the symmetry factor of the trivalent graph $i$. The part under the integral sign will be referred to as the integrand of the gauge theory amplitudes. 

Note the color factors still obey all possible color-Jacobi relations. Then there is a natural extension of color-kinematic duality to the integrand in two steps. First
\begin{suspicion}\label{susp:1}
For every loop level amplitude in a theory with adjoint matter only one can find numerators $n_j$ such that they satisfy the Jacobi relations of equation \eqref{eq:colorkinematicduality} and reproduce the integrand of the scattering amplitudes through \eqref{eq:BCJintegrand}.
\end{suspicion}
Second,
\begin{suspicion}\label{susp2}
Given a set of kinematic numerators which satisfy the kinematic Jacobi relations one can construct the integrand of a gravity amplitude $M$ by replacing the color factor in equation \eqref{eq:BCJintegrand} by another set of kinematic numerators
\begin{equation}
M^l_n=  \kappa^{n-2+2L}  \int \prod_{j=1}^l d^DL_j \sum_{\Gamma_i} \frac{1}{S_i} \frac{{n}_i \tilde{{n}}_i}{s_i}
\label{eq:BCJloopssquaring}
\end{equation}
\end{suspicion}
Assuming numerators $n_i$ can be found such that color kinematic duality \eqref{eq:colorkinematicduality} is satisfied unitarity implies that the double copy construction works as well, up to a subtlety identified below concerning generalized gauge transformations. This outline of a proof is again under the assumption that a set of color-dual numerators can be found which in addition has no spurious singularities.  In fact, the existence of such a set can be taken as an additional, stronger, conjecture. Both assumptions are at loop level still an open question. Several finite, low multiplicity examples for amplitudes have been reviewed in footnote 2. New results on color-kinematic duality for low-point form factors up to four loops have recently been obtained in \cite{Boels:2012ew}. See also \cite{toappearwithDonalRicardo} for the integrand of the finite one-loop amplitudes in pure Yang-Mills. 

Note that it is natural to enforce a similar choice of loop momentum routing for the integrands on both sides of \eqref{eq:BCJintegrand}. This convention will usually be understood throughout this article. At loop level the numerators are not unique either. To quantify this freedom, suppose as before they are shifted by some amount $\Delta_i$. In order for this shift to leave the amplitude invariant they have to fulfill the gauge condition
\begin{equation}\label{gaugeconditionintegrand}
\int \prod_{j=1}^l d^DL_j \sum_{\Gamma_i} \frac{1}{S_i} \frac{\Delta_i c_i}{s_i}= 0
\end{equation} 
The shifts can be made to satisfy the Jacobi relations as well. The gauge condition is not as specific as above since the integrand only has to satisfy
\begin{equation}
\sum_{\Gamma_i} \frac{1}{S_i} \frac{\Delta_i c_i}{s_i} = \textrm{integrates to zero}
\end{equation}
This leads to the question which terms vanish after integration. There are several ways terms can vanish after integration. The first is simple: the result might be a total derivative. An example of this is two terms that sum to zero after a shift in integration variable for one of them. I.e. 
\begin{equation}
\int dL^D \frac{1}{L^2 (L+p_1+p_2)^2} - \frac{1}{(L-p_1)^2 (L+p_2)^2} = 0
\end{equation}
Second, an integrand for a scattering amplitude can have no non-vanishing unitarity cuts. This is a standard reasoning to show this integrand integrates to zero. An example of this is the constant $1$, which famously integrates to zero in dimensional regularization,
\begin{equation}
 \int  d^DL \,\,1 = 0
\end{equation}
Vanishing of the cuts can be either purely algebraically or still involve a shift of the integration variable respecting the cut conditions. 

These varying ways of being able to satisfy the gauge condition \eqref{gaugeconditionintegrand} through vanishing terms have different implications for the squaring relations at the integrand level \eqref{gaugeconditionintegrand}. Where at tree level it is clear that all color-dual shifts $\Delta_i$ that obey equation \eqref{gaugecondition} will leave the squaring relation invariant, at loop level there is a problem with terms vanishing after integration since the numerators are generically momentum dependent. Hence the product of the gauge transformation with a kinematic numerators does not necessarily vanish. The example to keep in mind is the following type of integral,
\begin{equation}
\int_{-\infty}^{\infty} x dx =0 \quad \textrm{while} \quad \int_{-\infty}^{\infty} x^2 dx \neq 0
\end{equation}
The safe terms are those which vanish algebraically on cuts which fix the loop momenta completely. In this context it is useful to note that it was found in \cite{Bern:2009kd} in order to make color-kinematics work in explicit loop computations, terms have to be added to the Yang-Mills integrand whose color-factor vanishes and whose integrand integrates to zero as well. These terms turn out to be crucial in order to obtain the correct gravity amplitude using the double-copy construction. If they were excluded the gravity integrand would not have all correct unitarity cuts. A full discussion of this freedom is necessary and interesting but is also outside the scope of this article. If one adds the further restriction that \emph{local} numerators can always be found and these should be put into the double copy formula, then the ambiguity in the numerators is most probably to a large extent fixed, as it is in the known loop level examples. 

Recently, all-loop evidence for the conjecture was found investigating the IR structure of Yang-Mills and gravity \cite{Oxburgh:2012zr}. Some more evidence from a direct analysis of BCFW shifts for the integrand is contained in section \ref{sec:feynman}. In the following we will assume that color-dual kinematic numerators can be found at any loop order unless otherwise stated. Moreover, it will be assumed that for our results the subtlety in the role generalized gauge transformations in the double copy relations at loop level can safely be ignored. In particular, we assume that the double copy formula of equation \eqref{eq:BCJloopssquaring} can simply be applied in this form to the numerators which will be obtained below. 

\subsection{Solving Jacobi relations: the $\textrm{D}^3\textrm{M}$ basis}
To quantify the freedom in obtaining numerators in an equation like \eqref{eq:BCJtree} it is useful to solve all Jacobi relations between the color factors. This amounts to considering the total system of Jacobi relations as a massive system of linear equations of the type
\begin{equation}
A c = 0 \qquad A_{ij} \in \{-1,0,1\} \quad \forall \quad i,j
\end{equation}
with $c$ a $(2n-5)!!$ dimensional vector of color structures. The kernel of the matrix $A$ is a vector space which has several different baseses. One of these is particularly well-known \cite{DelDuca:1999rs} for the tree and one loop level and will be reviewed in this subsection. Obtaining an explicit solution to the system at higher loop orders is generically an open problem, but for most purposes of this article only the existence of a solution will be needed. 

\begin{figure}[h!]
\centering
\includegraphics[scale=0.5]{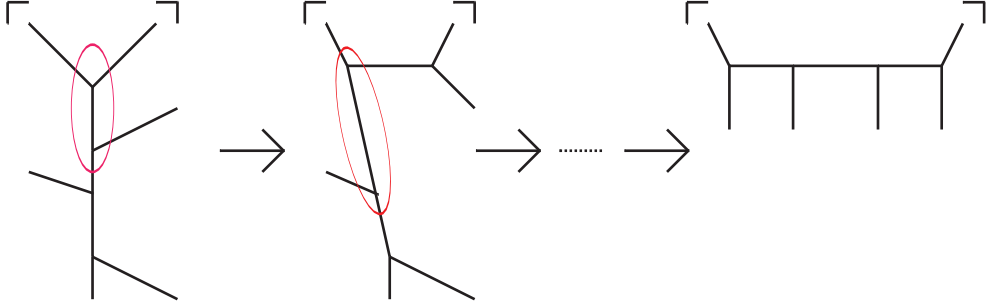}
\caption{\label{fig:maxjac}Schematic application of the Jacobi relations (red ellipsis) to maximize the distance between two singled out legs (denoted by the hat) in a given graph.}
\end{figure}

The system can be solved as follows: first single out two particles to be special, say $1$ and $n$. Then a solution to the Jacobi relations follows by using them to \emph{maximize} the number of structure constants encountered on the unique line through each trivalent graph from $1$ to $n$. For this maximization one simply draws the trivalent graphs as the line connecting particles $1$ and $n$ with all possible tree-graph subgraphs branching off from this line. These adornments can be `dissolved' into the connecting line using the Jacobi relation in graphical form of figure \ref{fig:IHX}. The result is a single line with all external particles attached directly (see figure \ref{fig:maxjac}). 

The resulting set is minimal since further application of any Jacobi identity would lead to non-trivial tree branches growing out of the connecting line. Hence the Jacobi relations can be solved in terms of the following set of structure constants,  
\begin{equation}\label{eq:minbasisI}
\{ f_{a_1,a_{\sigma(2)} \alpha_1} f^{\alpha_1}{}_{a_{\sigma(3)}\alpha_2} \ldots   f^{\alpha_{n-1}}{}_{a_{\sigma(n-1)} a_n} \} \qquad \forall \sigma \in  P\{2,\ldots, n-1\}
\end{equation}
where $P$ denotes the group of permutations of particles $\{2,\ldots, n\}$. For future purposes define 
\begin{equation}\label{fstrings}
F(1,2,3,\ldots, n) \equiv f_{a_1a_{2} \alpha_1} f^{\alpha_1}{}_{a_{3}\alpha_2} \ldots   f^{\alpha_{n-1}}{}_{a_{n-1} a_n}
\end{equation}
This set has $(n-2)!$ different elements: all permutations of particles $2$ through $n-1$. Scattering amplitudes can be expressed in this basis 
\begin{equation}\label{D3Mbasistree}
\mathcal{A}_n =g^{n-2}\sum_{\sigma \in  P\{2,\ldots, n-1\}} F(1,\sigma(2), \sigma(3), \ldots, \sigma(n-1), n)A_n^{\textrm{co}}(1,\sigma(2),\sigma(3),\ldots, \sigma(n-1), n) 
\end{equation}
where the coefficients $A^{\textrm{co}}$ were identified as the more well-known color-ordered amplitudes as was shown by Del Duca, Dixon, and Maltoni ($\textrm{D}^3\textrm{M}$) in \cite{DelDuca:1999ha, DelDuca:1999rs}. This is straightforward to prove in a modern way using on-shell recursion. As shown in \cite{DelDuca:1999rs}, the consistency of this equation with other possible choices of pairs of singled-out particles is a result of the Kleiss-Kuijf \cite{Kleiss:1988ne} relations for tree level color-ordered amplitudes. In fact, these relations can be seen as a consequence of the fact that a trace-based decomposition overcounts the number of independent group theory structures available at tree level. Note that a similar argument holds at one loop where a minimal basis obtained this way is a ring of structure constants \cite{DelDuca:1999rs}, while the corresponding relations for color-ordered amplitudes were found much earlier in \cite{Bern:1994zx}. The $\textrm{D}^3\textrm{M}$ basis at one loop for gluons, or more generally speaking particles in the adjoint, is explicitly given by
\begin{equation}\label{KKbasis1loop}
\mathcal{A}_n^{l=1}=g^n \sum_{\sigma \in  P\{1,\ldots, n\}/(Z_n\times Z_2)} \tilde{F}(\sigma(1),\sigma(2),\sigma(3),\ldots, \sigma(n))  \int \frac{d^Dl}{(2\pi)^D} I^{\textrm{co}}(\sigma(1),...,\sigma(n))
\end{equation}
where $\tilde{F}$ is the ring of structure constants (sometimes also called adjoint trace) given by
\begin{equation}
\tilde{F}(\sigma(1),\sigma(2),\sigma(3),\ldots, \sigma(n))\equiv f^{\alpha_1\sigma(a_1)}_{\qquad\;\;\;\alpha_2}f^{\alpha_2\sigma(a_2)}_{\qquad\;\;\; \alpha_3}...f^{\alpha_n\sigma(a_n)}_{\qquad\;\;\; \alpha_1}
\end{equation}
and  $I^{\textrm{co}}$ denoting the single trace color-ordered integrand. The sum above runs over the permutations of the $n$ external legs with inversions and reflections modded out. It is clear that in principle a solution to the Jacobi relations exists. An explicit solution would be interesting to find as this connects directly to extensions of the Kleiss-Kuijf relations to higher loop orders, see \cite{Naculich:2011ep, Edison:2011ta, Edison:2012fn} for work in this direction. An alternative solution to the Jacobi identities minimizing the distance between two given particles is presented in appendix \ref{app:mindist}.

\subsection{BCFW shifts and on-shell recursion}

In recent years on-shell recursion relations \cite{Britto:2005fq, Britto:2004ap} have been developed for gauge and gravity amplitudes at tree level and further steps to recursing loop level integrands and integrals have been taken, see  \cite{Feng:2011gc} for a review. The main power behind on-shell recursion is that it reconstructs an amplitude from its residues at (a subset of) its kinematic poles. To do so a complex parameter $z$ is introduced into the amplitude by shifting the momenta of any two legs of the amplitude, while keeping momentum conservation satisfied, by
\begin{equation}
\hat{p}_i= p_i + z\,q \qquad \hat{p}_j=p_j - z\,q
\end{equation}
The shift vector $q$ is chosen such that it keeps the masses of the legs invariant, i.e.
\begin{equation}
p_i\cdot q = p_j \cdot q = q\cdot q =0
\end{equation}
giving two complex solutions for $q$. The amplitude is now a function of a complex parameter $z$ and the original amplitude $A(0)$ can be recovered by a contour integral around the origin
\begin{equation}
A(0)=\frac{1}{2\pi i}\oint dz \frac{A(z)}{z}
\end{equation}
Under the assumption that $z=0$ is an isolated singularity the contour of integration can be extended to infinity to give (neglecting the prefactor)
\begin{equation}
A(0)=\oint dz \frac{A(z)}{z}=-\sum_{residues}\Big(\text{finite}\; z\Big)-\sum_{residue}\Big(z=\infty\Big)\label{BCFW}
\end{equation}
The residues at tree level for finite $z$ are products of lower point amplitudes summed over all internal states by unitarity. The residue at infinity lacks a similar physical interpretation, but if it can be shown to vanish \eqref{BCFW} constitutes an on-shell recursion relation: the right hand side only involves tree amplitudes with a lesser number of legs. Whether a theory obeys recursion relations consequently depends on the behavior of the amplitude for $z\rightarrow \infty$. If an amplitude falls off fast enough, i.e at least as $\sim(z^{-1})$, the residues at infinity are certainly absent and the amplitude can be computed through on-shell recursion. There has been work on establishing on-shell recursion even in the presence of boundary terms \cite{Feng:2009ei} and more recently there has been progress on on-shell recursion of Berends-Giele currents \cite{Britto:2012qi}.

The large-$z$ behavior of a Yang-Mills amplitude can be derived in several ways. All of them are variants of powercounting, i.e. tracing explicit powers of $z$ in Feynman diagrams. In doing so the shortest (simply connected) path between two shifted legs along which the $z$ dependence flows will be called ``hard line''. It can be shown that for a renormalizable gauge theory in $D \geq4$ dimensions \cite{ArkaniHamed:2008yf} that under a shift of two color-adjacent momenta a color-ordered amplitude scales as
\begin{equation}\label{eq:adscale}
\lim_{z\rightarrow \infty}A \sim \varepsilon^\mu(\hat{i}) \varepsilon^\nu(\widehat{i+1})\Big(z\eta_{\mu\nu}f_1(1/z)+z^0f_{2,\mu\nu}(1/z)+\mathcal{O}(1/z)\Big)
\end{equation}
where $f_i$ are polynomial functions in $(z^{-1})$ and $f_{2,\mu\nu}$ antisymmetric in its indices. The vectors $\varepsilon$ are the $z$-dependent polarization vectors of the shifted legs $i$ and $i+1$. 

The BCFW shift and on-shell recursion can be extended to the loop level at the level of the integrand \cite{ArkaniHamed:2010kv, Boels:2010nw}. For the shift the above result in gauge theory follows from evaluating the amplitude through its Feynman graphs in the natural $q_{\mu} A^{\mu} = 0$ space cone gauge. The result for the shift for loop integrands is unchanged from the tree level result, i.e. it is given by \eqref{eq:adscale}. For integrands the notion of hard line follows from a particular choice for the routing of loop momenta.

\subsubsection*{BCFW scaling analysis of gravity amplitudes at tree level and beyond}
The emergence of on-shell recursion relations for gauge theory amplitudes raised the natural question if similar constructions also hold for gravity theories. Here the outlook was originally very negative: simple powercounting of the involved Feynman graphs for BCFW shifts gives for $n$ gravitons a $z^{n-2}$ divergence at $z \rightarrow \infty$. Study of examples in \cite{Bedford:2005yy} and \cite{Cachazo:2005ca} showed, however, that the recursion relations could actually hold, which was subsequently proven in \cite{Benincasa:2007qj} in four and \cite{ArkaniHamed:2008yf} in $D$ dimensions. See \cite{Cheung:2008dn} for the extension to matter couplings. These works showed the actual divergence is a double copy of equation \eqref{eq:adscale}: the gravity polarization tensors can simply be decomposed in polarization tensors of two copies of gauge theory. A particular lucid argument for scaling of gravity amplitudes using a background field technique appeared in \cite{ArkaniHamed:2008yf} where the scaling of an Einstein gravity amplitude was proven to be
\begin{multline}
 \lim_{z\rightarrow \infty} M(z) \sim\left[ \varepsilon^\mu(\hat{i})   \varepsilon^{\tilde{\mu}}(\hat{i})\right]\left[ \varepsilon^\nu(\widehat{i+1})   \varepsilon^{\tilde{\nu}}(\widehat{i+1}) \right] \Big(z\eta_{\mu\nu}f_1(1/z)+z^0f_{2,\mu\nu}(1/z)+\mathcal{O}(1/z)\Big) \\
\Big(z\eta_{\tilde{\mu}\tilde{\nu}} f_1(1/z)+z^0f_{2,\tilde{\mu}\tilde{\nu}}(1/z)+\mathcal{O}(1/z)\Big)
\label{arkanigrav}
\end{multline}
where $B$ is as before an antisymmetric tensor. The product of polarization tensors in brackets belongs to the graviton, dilaton and two form.  In passing we note that a form of the gravity tree level amplitudes which demonstrates fairly explicitly this scaling behavior has recently been found in four dimensions \cite{Cachazo:2012kg, Cachazo:2012pz}. 

This much-improved behavior under BCFW shifts at tree level compared to powercounting Feynman graphs immediately raises the question by what physical symmetry this is driven. Furthermore, based on experience with gauge theory integrands it is natural to investigate if the results obtained for tree level amplitudes can be carried over to the integrand level. It will be shown below, up to some assumptions, that the driving force of the cancellations in gravity is color-kinematic duality. Moreover, it will be argued the shift of the gravity integrand is indeed the same as the tree amplitude implying extensive cancellations taking place within the perturbative sum.

\subsection{Generalized inverses}\label{sec:geninverses}

The concept of generalized inverses will be central in our study of BCFW shifts of color-dual kinematic numerators below. This is a well-developed mathematical subject of which some elements will be briefly reviewed here. See e.g. \cite{Benisrael80} for a more complete introduction. 

First generalized inverses for matrices are discussed. The purpose of introducing matrices is to solve systems of linear equations,
\begin{equation}
A x = b
\end{equation}
Here $A$ generically maps a vector space $V$ to itself. If $A$ is a square matrix of rank $n$ it is well known that a unique inverse $A^{-1}$ exists defined by
\begin{equation}
A^{-1} \, A\,=\,I\,=\,A\, A^{-1}
\end{equation}
which can be used to solve the equation to give
\begin{equation}
x = A^{-1} b
\end{equation}

In the case of singular or non-square matrices an inverse in this sense does not exist. However, the system of equations given by $A x = b$ can still have solutions: it is generically under- or overdetermined. This realization then leads to the theory of generalized inverses of matrices as was introduced first by Moore and later independently Penrose  \cite{Moore1920, Penrose:1955vy}. Their work generalizes the notion of a matrix inverse to the cases of singular and rectangular matrices, hence called `generalized inverse'. The generalized inverse $A^+$ of a $n\times m$ matrix $A$ is a $m \times n$ matrix defined here by the following condition only,
\begin{equation}\label{consitencygeninv}
A\, A^+ \,A=A
\end{equation}
which reduces to the $A^{+}=A^{-1}$ if A is square and of full rank. In the more general case where $A$ is singular or rectangular there is a solution to the system of equations
\begin{equation} \label{eq:axisb}
A\,x=b
\end{equation} 
only if consistency conditions hold. For instance, multiplying left hand side with a transpose vector $c^T$ for which $c^T A =0$ holds should give $c^T b = 0$. If $A$ is a matrix these conditions can be shown to be equivalent to
\begin{equation}\label{linsysconsistency}
A\,A^+ b=b
\end{equation}
This equation is certainly necessary as can be checked by multiplying left and right hand side in equation \eqref{eq:axisb} by $A A^+$. If the consistency conditions hold then the most general solution to the linear system is given by
\begin{equation}\label{eq:geninvsol}
x=A^+ \,b+(I-A^+\,A)\,y
\end{equation}
with $y$ an arbitrary $n$ dimensional vector and $I$ is the appropriate $n\times n$ identity. It is easy to check that $(I-A^+A)y$ is in the kernel of $A$. Less obvious is that it spans the kernel,
\begin{equation}
\ker A = \{(I-A^+A)y \, | \, y \in \mathbb{C}_n \}
\end{equation}
Moreover, it is straightforward to check that \eqref{eq:geninvsol} is a solution to the linear system since
\begin{equation}
A\,x=A\,(A^+\,b+(I-A^+\,A)\,y)=A\,A^+b=b
\end{equation}
The generalized inverse $A^+$ specified by equation \eqref{consitencygeninv}  is not unique as it can be changed by transformations which leave equation \eqref{consitencygeninv}  invariant. More precisely, if $A^+$ satisfies \eqref{consitencygeninv} then 
\begin{equation}\label{eq:othergeninverses}
(A^+)' = A^+ + (I- A^+\, A)\, Y + W \,(I- A \,A^+)
\end{equation}
also satisfies equation \eqref{consitencygeninv} for arbitrary matrices $Y$ and $W$ of size $m\times m$ and $n\times n$ respectively. Note that the extra pieces are related to the kernel of $F$. Further note that the generalized inverse \eqref{consitencygeninv}
 is not unique in the sense that one can demand $A^+$ to obey further properties like reality conditions, see for instance appendix \ref{sec:pseudoinv} or \cite{Benisrael80} for further examples and information regarding the different types of generalized inverses.

\subsubsection*{Example of a generalized inverse}
As an instructive example consider a singular matrix $A$ given by
\begin{equation}
A=\begin{pmatrix} 1 & 0 \\ 0 &0 \end{pmatrix}
\end{equation} 
A generalized inverse $A^+$ of this matrix is given by
\begin{equation}
A^+=\begin{pmatrix} 1 & 0 \\ 0 &0 \end{pmatrix}
\end{equation}
as it obviously satisfies \eqref{consitencygeninv}. The matrix problem
\begin{equation}
A x = b
\end{equation}
can only have a solution in this case if
\begin{equation}\label{eq:defofbeta}
A A^+ b = b \quad \rightarrow \quad b = \left( \begin{array}{c} \beta \\ 0  \end{array}\right)
\end{equation}
for some constant $\beta$. The matrix $I  -A ^+A$ is simply
\begin{equation}
I  -A^+ A = \left(\begin{array}{cc} 0 & 0 \\ 0 & 1\end{array}\right)
\end{equation}
and its easy to see that 
\begin{equation}
\ker A = \{(I-A^+A)y \, | \, y \in \mathbb{C}_n \}
\end{equation}
indeed spans the kernel of A. Hence the most general solution to $Ax=b$ of the form
\begin{equation}
x = A^+ b + (I  -A^+ A) w
\end{equation}
reads in this example
\begin{equation}
x =  \left( \begin{array}{c} \beta \\ w  \end{array}\right)
\end{equation}
where $\beta$ was defined in equation \eqref{eq:defofbeta} and $w$ is arbitrary. The reader is encouraged to experiment with some alternative generalized inverses in this example generated by equation \eqref{eq:othergeninverses}.

\subsubsection*{Extension to more general linear maps}

In the above exposition a main assumption is that the matrix $A$ is a linear map between two vector space $V_1$, $V_2$ over some field $K$. The concept of generalized inverse however can be generalized beyond this to linear maps not necessarily on vector space over field. In category theory terms, such mappings are called morphism in the category of modules over a given ring.

To get an idea of what is involved, consider the ring of integers $V_1 = \mathbb{Z}$ and the ring of integers modulo 2, $V_2 = \mathbb{Z} \mod 2$. Let $A$ be a linear map,
\begin{equation}
A: V_1 \rightarrow V_2
\end{equation}
given by multiplication by $3$,
\begin{equation}
A(x) = 3 x
\end{equation}
Now consider the equation 
\begin{equation}
A(x) \sim b
\end{equation}
Here the similarity sign is used to stress that this is an equation in the space $V_2$. By construction, $b$ can only take the values  $0$ or $1$. In this case, it is easy to construct solutions directly for $b$ either $0$ or $1$:
\begin{equation}
b\sim0 : x = 2 k \quad \textrm{and if }  b\sim1: x= 2 k +1  \qquad \textrm{for } k \in \mathbb{Z}
\end{equation}
It is not hard to recognize the kernel of the linear map $A$,
\begin{equation}
\textrm{ker}A = 2 k \ \textrm{for } k \in \mathbb{Z}
\end{equation}
The generalized inverse $A^+$ in this case is a map $V_2 \rightarrow V_1$ such that 
\begin{equation}\label{eq:defgeninvexamp}
A A^+ A(x) \sim  A (x) \qquad \forall x \in \mathbb{Z} 
\end{equation}
In this case, any $1$-to-$1$ map in the space $V_2$ can be used as the generalized inverse. In this example the map $A$ is invertible: there is a solution for every $b$. However, for the setup under study this is not always true. Consider
\begin{equation}
C: V_1 \rightarrow V_2 \qquad | \qquad  C(x) = 2 x
\end{equation}
It is obvious that an equation of the type 
\begin{equation}
C(x) \sim b
\end{equation}
for this map only has a solution if $b\sim 0$. For the generalized inverse again any  $1$-to-$1$ map in the space $V_2$ will do to satisfy equation \eqref{eq:defgeninvexamp}. Note that also a map which maps any number unto $0$ will work. A consistency condition of the equation obtained by acting on left and right hand sides with $C C^+$ is 
\begin{equation}
C C^+ (b) \sim b
\end{equation}
which is indeed only satisfied if $b\sim 0$. In this case this equation is apart from necessary also sufficient. The most general solution for the equation $C(x) \sim b$ is
\begin{equation}
x = C^+ b + \textrm{ker C}
\end{equation}

In the example the kernel of the map $A$ arises as the two spaces $V_1$ and $V_2$ have a different intrinsic dimensionality: the linear maps involved are in general many to one. This changes the setup slightly compared to the matrix case. Let $A$ be a general map between vector spaces $V_1$ and $V_2$. Then the generalized inverse of $A$ is the linear map $A^+$ for which 
\begin{equation}\label{eq:defgeninvmaps}
A A^+ A(x) \sim  A (x) \qquad \forall x \in V_1 
\end{equation}
for an equivalence relation on the space $V_2$. The equation 
\begin{equation}\label{eq:tobesolvedequation}
A(x) = b
\end{equation} 
has in general a solution only if 
\begin{equation}
A A^+ (b) \sim b
\end{equation}
In general this is not sufficient as the maps involved are in general many to one. The more general consistency condition is that for every linear map $C: V_2 \rightarrow V_2$ such that 
\begin{equation}
C (A x) \sim 0 \Rightarrow C (b) \sim 0
\end{equation}
A solution of equation \eqref{eq:tobesolvedequation} is given in terms of the generalized inverse by 
\begin{equation}\label{eq:geninvmaps}
x = A^+ b + \textrm{ker} A
\end{equation}
As before, the generalized inverse is not unique and can be modified by kernel mappings. More explicitly, if $A^+$ satisfies equation \eqref{eq:defgeninvmaps}, then so does
\begin{equation}
A^+ \rightarrow A^+ + C_1 + C_2
\end{equation}
where $C_i$ are maps $V_2 \rightarrow V_1$ such that
\begin{equation}
A C_1 x \sim 0 \qquad \forall x \in V_2
\end{equation}
and
\begin{equation}
C_2 A x \sim 0 \qquad \forall x \in V_1
\end{equation}
hold. This is the linear map analog of equation \eqref{eq:othergeninverses}.


\section{BCFW shifts of gravity amplitudes from gauge theory}\label{sec:amplitudes}

In this section the following question will be studied: can color-kinematic duality be used to derive BCFW shifts of gravity tree level amplitudes from gauge theory amplitudes? The main tool is the realization that the complete first step in the color-kinematic duality can be rephrased as a problem in linear algebra. This will lead to a singular system of linear equations which can be inverted in the generalized sense as introduced above to give the kinematic numerators $n$ in terms of color-ordered amplitudes. By BCFW-shifting the large-$z$ scaling of the numerators can then be read off and by using the double copy construction the large-$z$ behavior of the gravity tree amplitudes will be obtained. In this section the kinematic numerators are always assumed to be color-dual.

 \subsection{Warmup: four-point amplitudes revisited}\label{sec:4amplitudes}
Let us illustrate the strategy at four points. The Yang-Mills four-point amplitude in the cubic BCJ representation is given by
\begin{equation}
 \mathcal{A}_4=\frac{n_s c_s}{s}+\frac{n_t c_t}{t}+\frac{n_u c_u}{u}
\end{equation}
where $s$, $t$, and $u$ are the Mandelstam variables at four points and $c_i$ the color factors given by
\begin{equation}
 c_s=f^{a_1a_2}_{\;\;\;\quad b}f^{ba_3a_4} \quad c_t=f^{a_2a_3}_{\;\;\;\quad b}f^{ba_4a_1} \quad c_u=f^{a_1a_3}_{\;\;\;\quad b}f^{ba_4a_2}
\end{equation}
The $n_i$ denote the corresponding kinematic numerators. Both, color factors and kinematic numerators satisfy the same Jacobi relation, i.e.
\begin{equation}
 c_u=c_s-c_t \quad \text{and} \quad  n_u=n_s-n_t
\end{equation}
Alternatively, the four-point amplitude can be written in terms of the $\textrm{D}^3\textrm{M}$ basis \eqref{D3Mbasistree} singling out legs $1$ and $4$
\begin{equation}
 \mathcal{A}_4=c_s A(1234)+c_u A(1324)
\end{equation}
where $A$ denotes color-ordered amplitudes. As these two representations are equivalent one obviously has
\begin{equation}
\frac{n_s c_s}{s}+\frac{n_t c_t}{t}+\frac{n_u c_u}{u}\equiv c_s A(1234)+c_u A(1324)
\end{equation}
Inserting the Jacobi relations for color factors and kinematic numerators one the left hand side can be used to eliminate $n_t$ and $c_t$ and arrives at
\begin{equation}
 c_s\Big(\frac{n_s}{s}+\frac{n_s-n_u}{t}\Big)+c_u\Big(\frac{n_u}{u}-\frac{n_s-n_u}{t}\Big)=c_s A(1234)+c_u A(1324)
\end{equation}
which can nicely be written as a matrix equation
\begin{equation}\label{numeratorsystem}
 c^iF_{ij}n^j=c^i A_i \quad \Rightarrow \quad F_{ij}n^j= A_i
\end{equation}
with
\begin{equation}\label{F4pts}
 F=\begin{pmatrix}
       \frac{1}{s}+\frac{1}{t} & -\frac{1}{t}\\
       -\frac{1}{t} & \frac{1}{t}+\frac{1}{u}\\
      \end{pmatrix}\quad n=\begin{pmatrix}n_s \\ n_u \end{pmatrix} \quad  A=\begin{pmatrix}A(1234) \\ A(1324) \end{pmatrix}
\end{equation}
$F$ is symmetric in its indices. Its determinant vanishes on-shell by momentum conservation
\begin{equation}
det(F)=\frac{s+t+u}{stu}=0
\end{equation}
and has rank unity. For this reason the previously introduced concept of a generalized inverse (compare section \ref{sec:geninverses}) has to be applied to invert \eqref{numeratorsystem}. One particularly simple representation is given by
\begin{equation}
 F^+=\begin{pmatrix}0 & 0\\0&\frac{t(s+t)}{s}      \end{pmatrix}
\end{equation}
For a solution to the linear problem in equation \eqref{numeratorsystem} to exist the color ordered amplitudes have to satisfy
\begin{equation}\label{consistencycond}
 FF^+ A = A
\end{equation}
Spelling this out (using momentum conservation $s+t+u=0$) one finds
\begin{equation}
\begin{pmatrix} \frac{1}{s}\Big(A(1234)-A(1324)\frac{u}{s}\Big)+A(1324)\frac{u}{s} \\ A(1324)+\frac{1}{u}\Big(A(1234)-A(1324)\frac{u}{s}\Big) \end{pmatrix} \stackrel{!}{=}\begin{pmatrix}A(1234)\\ A(1324) \end{pmatrix}
\end{equation}
This boils down to 
\begin{equation}
A(1234)-A(1324)\frac{u}{s}=0
\end{equation}
which is nothing but the BCJ relation at four points. In other words, in this four point example the BCJ relations are a necessary and sufficient condition that the kinematic numerators satisfying the Jacobi relation exist. The most general solution to equation \eqref{numeratorsystem} is given by
\begin{equation}
\left(\begin{array}{c} n_s \\ n_u \end{array} \right) = \begin{pmatrix}0 & 0\\0&\frac{t(s+t)}{s} \end{pmatrix} \begin{pmatrix}A(1234) \\ A(1324) \end{pmatrix} +  \begin{pmatrix}1 & 0\\ -\frac{u}{s}&0 \end{pmatrix} \begin{pmatrix}w_1 \\ w_2\end{pmatrix} 
\end{equation}
for some vector $\vec{w}$. The second term involving the vector $\vec{w}$ spans the kernel of $F$. Note that this kernel has a physical interpretation: it is the space of (color-dual) generalized gauge transformations. This can be demonstrated by evaluating equation \eqref{gaugecondition} using the solution to the Jacobi identities employed here. 

The behavior of the color-dual kinematic numerators under BCFW shifts can now be investigated in the four particle case. It is natural to shift particles $1$ and $4$, i.e. those kept fixed in the $\textrm{D}^3\textrm{M}$ basis. Under this shift the matrix entries of $F$ in \eqref{F4pts} scale as $z^0+\mathcal{O}(1/z)$. The generalized inverse $F^+$ scales as
\begin{equation}
\begin{split}
\lim_{z\rightarrow \infty}F^+\sim&
      \begin{pmatrix}0&0\\0&t+\mathcal{O}(1/z)     \end{pmatrix} \sim z^0  \begin{pmatrix}0&0\\0&1 \end{pmatrix}+\mathcal{O}(1/  z)
\end{split}
\end{equation}
while
\begin{equation}
\lim_{z\rightarrow \infty} \left(I-F^+ F\right) w \sim  \begin{pmatrix}1 & 0\\ 1&0 \end{pmatrix} \begin{pmatrix}w_1 \\ w_2\end{pmatrix} +  \mathcal{O}(1/  z)
\end{equation}
Finally, the large-$z$ scaling of the numerators is given by from \eqref{eq:geninvsol} as
\begin{equation}
 \lim_{z\rightarrow \infty} n_i=\lim_{z\rightarrow \infty}\Big( F^+(z) \,A(z)+ker(F)(z) \Big)_i \sim \lim_{z\rightarrow \infty} A_i(z) +\mathcal{O}\Big(\frac{A_i(z)}{z}\Big) + \textrm{Kernel}
 \end{equation}
Note that the kernel vanishes in \eqref{numeratorsystem} and does not contribute to the gravity amplitude in the double copy formula. That means that up to gauge transformations the numerators of the four point YM amplitude scale like color-ordered amplitudes adjacently shifted. The scaling of these objects is known from equation \eqref{eq:adscale} and so the scaling of the kinematic numerators is given by
\begin{equation}
 \lim_{z\rightarrow \infty} n_i\sim   \varepsilon(\hat{1})_\mu\varepsilon(\hat{4})_\nu \Big(z \eta^{\mu\nu}f_i (1/z)+z^0 B_i^{\mu\nu}\Big)+\mathcal{O}(1/z)
\end{equation}
where $\varepsilon$ are the gluon polarization vectors, $f_i(1/z)$ is a function in $1/z$ and $B_i^{\mu\nu}$ is an antisymmetric matrix. By the double copy formula in equation \eqref{eq:squaringrelation} this behavior of the numerators up to generalized gauge transformations can be squared to give the shift of the corresponding gravity amplitude in $\mathcal{N}=0$ SUGRA,
\begin{multline}
\lim_{z\rightarrow \infty} M_4 \sim \varepsilon(\hat{1})_\mu\tilde{\varepsilon}(\hat{1})_{\kappa} \varepsilon(\hat{4})_\nu \tilde{\varepsilon}(\hat{4})_{\lambda}  \\ 
 \Big(z \eta^{\mu\nu}f(1/z)+z^0 B^{\mu\nu}+\mathcal{O}(1/z)  \Big) \Big(z \eta^{\kappa\lambda}f(1/z)+z^0 B^{\kappa\lambda}+\mathcal{O}(1/z)\Big) 
\end{multline}
The reasoning of this subsection will be extended in this section to all multiplicity at tree level and in the next section to the integrand. Before doing so the construction of numerators through linear algebra will be highlighted as this might have wider applications.


\subsection{Rephrasing tree level color-kinematic duality as linear algebra}

The first step is to solve the system of Jacobi relations. That is, there exists a basis for the color factors $\breve{c}^j$ which in the tree level case is known to have dimension $(n-2)!$. Moreover, any color factor of a tree level trivalent connected graph can be expressed as a linear combination of these, i.e.  a rectangular matrix $W_{\underline{i}\,j}$ exists for which
\begin{equation}
c_{\underline{i}} = W_{\underline{i}\,j} c^j_b \qquad j = 1, \ldots, (n-2)! \qquad \underline{i} = 1, \ldots, (2n -5)!!
\end{equation}
At tree level, the $\textrm{D}^3\textrm{M}$ basis was reviewed above as an example of such a basis. Since this is just solving Jacobi relations this solution can be used in equation \eqref{eq:BCJtree} for both color as well as kinematic part. This leads to 
\begin{equation}
\mathcal{A}_n= g_{ym}^{n-2} \sum_{j,k} \breve{n}^j \left(\sum_{\Gamma_i}\frac{W_{ij} W_{ik}}{s_i}\right) \breve{c}^k
\end{equation}
which in turn leads to 
\begin{equation}
\mathcal{A}_n= g_{ym}^{n-2} \sum_{j,k} \breve{n}^j F_{kj} \breve{c}^k
\end{equation}
with the symmetric $(n-2)! \times (n-2)!$ matrix $F$
\begin{equation}
F_{jk} \equiv  \left(\sum_{\Gamma_{\underline{i}}}\frac{W_{\underline{i} \,j} W_{\underline{i}\,k}}{s_{\underline{i}}}\right) 
\end{equation}
Moreover, the amplitude itself can be expressed in the color basis, 
\begin{equation}
\mathcal{A}_n = \sum_j {A}_{n,j} \breve{c}^j
\end{equation}
Hence if color-dual numerators exist then 
\begin{equation}\label{eq:findingnumerators}
F_{jk}  \breve{n}^k  = {A}_{n,j} 
\end{equation}
must hold for a symmetric matrix $F$. Note this equation (but not the inverse discussed below) also appears in \cite{Vaman:2010ez}. Hence if $F$ were invertible it would be proven numerators always exist and that they are unique. 

However, as shown above in the four point example the matrix $F$ is in general singular. Moreover, we strongly suspect but have not been able to prove in general that the kernel of $F$ is exactly the set of relations generated by the fundamental BCJ relations. This is basically a conjecture formulated in different ways in \cite{Bern:2008qj}, see also \cite{Vaman:2010ez}. Its essence is that the BCJ relations are the necessary and sufficient conditions for a color-dual representation to exist for scattering amplitudes.  One can reason as follows: since $F$ is symmetric, every vector in the null-space generates a relation for the vector of color ordered amplitudes by contracting left and right had side of the above equation. Using the $\textrm{D}^3\textrm{M}$ basis the set of amplitudes on the right hand side is the set of color-ordered amplitudes with two particles adjacently ordered. The most general set of relations for these color ordered tree amplitudes known and proven through other means is the collection of BCJ relations. Vice versa, every BCJ relation is a null vector of $F$. This makes the dimension of the kernel of the matrix $F$ at least $(n-2)!-(n-3)!$ and the rank of $F$ maximally $(n-3)!$ under the assumption color-kinematic duality holds. We have checked this explicitly up to six points. Note that the kernel of $F$ is the set of (color-dual) generalized gauge transformations, as noted above in the four particle example. 

The matrix problem in equation \eqref{eq:findingnumerators} can be solved in terms of a generalized inverse $F^+$ if and only if the consistency condition
\begin{equation}\label{eq:conditiononamps}
F F^+  {A} =  {A}
\end{equation}
holds. As argued above, these conditions should be equivalent to the BCJ relations. Moreover, these conditions are equivalent to the existence of a set of Jacobi-satisfying numerators at tree level, i.e. theorem \ref{theo:BCJpart1tree}. As there are examples of such sets, the conditions must hold. Hence the system \eqref{eq:findingnumerators} can be solved to give
\begin{equation}
 \breve{n}  = F^{+} {A}_{n} + (I- F^+ F) v
\end{equation}
for an arbitrary $(n-2)!$-dimensional vector $v$. Plugging this solution into the double copy formula in theorem \ref{theo:BCJpart2tree} and taking into account \eqref{eq:geninvmaps} yields the most general form for the gravity amplitude in this language
\begin{equation}\label{eq:BCJtreeonsteroids}
\mathcal{M}_n= \kappa^{n-2} (n_b)^T F \tilde{n}_b = \kappa^{n-2}   ({A}_{n})^T \left(F^+ + (I- F^+\, F)\, Y + W \,(I- F \,F^+) \right) \tilde{{A}}_{n} 
\end{equation}
as the natural consequence of the double copy construction in equation \eqref{eq:squaringrelation}. Since the amplitudes satisfy the consistency condition this can for practical purposes by shortened to
\begin{equation}
\mathcal{M}_n= \kappa^{n-2}   ({A}_{n})^T \left(F^+ \right) \tilde{{A}}_{n} 
\end{equation}
Note that one particular example of a matrix $F^+$ can be read off from this equation by comparing to the KLT relation obtained in \cite{BjerrumBohr:2010ta}. In general it would be interesting to further explore the relation of the linear algebra approach to the momentum kernel of \cite{BjerrumBohr:2010hn} especially through its link to string theory.

Note that this analysis shows that most features of color-kinematic duality are captured by the features of the matrix $F$ and its generalized inverse. The properties of these matrices are largely independent of the intricacies of Yang-Mills or gravity theories: they arise in \emph{any} theory which has trivalent vertices made out of two Jacobi-satisfying structure constants. Other examples of these have been studied before  \cite{Monteiro:2011pc, BjerrumBohr:2012mg}. A particular class of these are scalar field theories, constructed to consist of a single massless scalar and only a three vertex which consists of the product of two structure constants satisfying the Jacobi relations. In other words, its perturbative series is designed to yield equation \eqref{eq:BCJtree} directly. Note that this class involves specifying two `gauge groups' and is therefore infinite dimensional.  This class of quantum field theories will be referred to as `trivalent scalar theories'.

\subsection{BCFW shifts of gravity amplitudes constructed by double copy} 

The general setup just given can now be used to generalize the results at four points to arbitrary points at tree level. For concreteness use the $\textrm{D}^3\textrm{M}$ basis fixing legs $1$ and $n$, to set up the problem,
\begin{equation}\label{eq:treelevelnpointsD3m}
 \sum_{\Gamma_i}\frac{n_i{c}_i}{s_i}=\sum_{\sigma\in S_{n-2}} \breve{c}_\sigma A(1,\sigma(2),...,\sigma(n-1),n) \equiv \sum_{i} \breve{c}_i A_i
 \end{equation}
where $\Gamma_i$ runs over all cubic graphs and $S_{n-2}$ runs over all permutations of $(2,...,n-1)$. As demonstrated above, this reduces to
\begin{equation}
F_{ij} \breve{n}^j=   A_i
 \end{equation}
which can then be solved in terms of a generalized inverse $F^+$. As argued above, this equation is consistent. 

To obtain the large-$z$ scaling of the numerators a BCFW-shift of particles $1$ and $n$ will be applied. It is clear how $F$ scales under BCFW shifts: as $\sim z^0$. However, this does not automatically imply in general $F^+$ scales as $z^0$. The canonical counter-example is
\begin{equation}
A = \left(\begin{array}{ccc} 1 & 0 & 0 \\ 0 & \epsilon & 0 \\ 0& 0 & 0\end{array} \right) \qquad A^+ =\left( \begin{array}{ccc} 1 & 0 & 0\\ 0 & \frac{1}{\epsilon} & 0\\ 0& 0 & 0 \end{array} \right)
\end{equation}

To obtain the scaling of the generalized inverse $F^+$ under large BCFW shifts it will be useful to express the Yang-Mills amplitude via a basis of non-trivial eigenvectors of $F$. The key to doing this is the observation mentioned above that amplitudes in the class of trivalent scalar theories satisfy the consistency conditions \eqref{eq:conditiononamps} and therefore span the non-trivial eigenvectors of $F$. As argued above, the dimension of this space is $(n-3)!$. Therefore for sufficiently rich choices of scalar field theories any solution to equation \eqref{eq:conditiononamps} can be expanded in terms of them, up to generalized gauge transformations, 
\begin{equation}\label{eq:expandingymnumsinscalarnums}
\breve{n}_{ym,i}= \sum_{\bar{K}}^{(n-3)!} \alpha_{\bar{K}} (\breve{n}_i^{\bar{K}})
\end{equation}
where $n_{ym}$ are the Yang-Mills numerators, $\breve{n}_i^{\bar{K}}$ the trivalent scalar numerators, and $\alpha_{\bar{K}}$ expansion parameters. The numerators on the right hand side must \emph{not} be in the kernel of $F$,
\begin{equation}
F_{ij} \breve{n}^{j,\bar{K}} \neq 0
\end{equation}
and be linearly independent. Equation \eqref{eq:expandingymnumsinscalarnums} is equivalent to an observation made in \cite{BjerrumBohr:2012mg}. Now it is known how $F$ acts on the combination in equation \eqref{eq:expandingymnumsinscalarnums} through equation \eqref{eq:findingnumerators},
\begin{equation}\label{eq:expaamplitsscalarbas}
A_{j,YM} = F_{ji} n^i_{ym}= \sum_{\bar{K}}^{(n-3)!} \alpha_{\bar{K}} (\Theta_j^{\bar{K}}) 
\end{equation}
where $\Theta^{\bar{K}}$ are the amplitudes in the choice of trivalent scalar theories labelled by $\bar{K}$. The construction can also be set up in the other direction, starting with the amplitudes. The point of this expansion here is that for the trivalent scalar theories, 
\begin{equation}
 \breve{n}^{\bar{K}}  = F^{+} \Theta^{\bar{K}} + (I- F^+ F) v^{\bar{K}}
\end{equation}
holds. In the class of trivalent scalar theories, it can now be shown that $F^+$ scales as $\sim z^0$, up to a transformation in the kernel of $F$, i.e. a generalized gauge transformation. That is, the large-$z$ scaling of $F$ and $F^+$ is given by
\begin{equation}
\begin{split}
&\lim_{z\rightarrow \infty} F = z^0+\mathcal{O}(1/z)\\ 
&\lim_{z\rightarrow \infty} F^+ = z^0+\mathcal{O}(1/z) \quad \text{up to terms from\;}ker(F)
\end{split}
\end{equation}
To show this, consider the $\textrm{D}^3\textrm{M}$ basis at tree level which singles out the BCFW shifted particles $1,n$. All scalar field theory amplitude coefficients in this basis scale as $z^0$ for this shift. The numerators in the trivalent scalar theories manifestly scale as $z^0$, up to generalized gauge transformation. Hence $F^+$ must scale as $z^0$.

From equation \eqref{eq:expaamplitsscalarbas} it therefore follows that for Yang-Mills theories the essential information about scaling behavior of the numerators is captured by the coefficients $\alpha$.  In the case at hand the $\textrm{D}^3\textrm{M}$ basis is employed that singles out particles $1$ and $n$. Shifting these amounts to a color adjacent shift  for each component of the left hand side of equation \eqref{eq:expaamplitsscalarbas}. Since the scalar field theories manifestly scale as $z^0$, this shows that the scaling of the coefficients $\alpha$ follows immediately from the BCFW shift of color-adjacent gluons shown in equation \eqref{eq:adscale}
\begin{equation}
  \lim_{z\rightarrow \infty}  \alpha_{\bar{K}} \sim \varepsilon(\hat{1})_\mu\varepsilon(\hat{n})_\nu \Big(z \eta^{\mu\nu}f_{\bar{K}} (1/z)+z^0 B_{\bar{K}} ^{\mu\nu}\Big) +\mathcal{O}(1/z)
\end{equation}
Plugging this into equation \eqref{eq:expandingymnumsinscalarnums} yields the result that the large-$z$ behavior of the kinematic color-dual Yang-Mills tree numerators at arbitrary multiplicity is (suppressing the subscript)
\begin{equation}\label{numtreelevel}
 \lim_{z\rightarrow\infty} n_i \sim \lim_{z\rightarrow \infty} \varepsilon(\hat{1})_\mu\varepsilon(\hat{n})_\nu \Big(z \eta^{\mu\nu}f(1/z)+z^0B^{\mu\nu}\Big)_i+\mathcal{O}(1/z) + \textrm{kernel}
\end{equation}
up to generalized gauge transformations.  In writing this result we have used the fact that all numerators $n_i$ can be expressed as linear combinations of the basis numerators $\breve{n}$. These linear combinations only involve numbers. Note that strictly speaking this is an upper bound on the scaling: it could scale better. The result of equation \eqref{numtreelevel} is sufficient to obtain the BCFW shift of the gravity amplitude through equation \eqref{eq:squaringrelation}.

It follows immediately from the numerators or equivalently  from equation \eqref{eq:BCJtreeonsteroids} that the (extended) Einstein gravity tree level amplitude at arbitrary multiplicity scales as two copies of Yang-Mills as in equation \eqref{arkanigrav}. This result itself was obtained through the background field method in \cite{ArkaniHamed:2008yf}. What is new here is the explicit proof that the mechanism behind the large suppression in $z$ scaling compared to the na\"ive powercounting result  $\sim z^{n-2}$ is color-kinematic duality.

\subsection*{Inclusion of renormalizable matter}

So far only gluonic matter has been considered and the question can be asked what happens if one adds fermionic and / or scalar matter in the adjoint representation to the gluons. The answer is simple: the only point were the field content played a role in the above argument is the shift of the tree level Yang-Mills amplitude in the $\textrm{D}^3\textrm{M}$ basis. Now from the generic arguments above it follows that this basis also exists for quite generic fermionic or scalar matter transforming in the adjoint of the gauge group. This and the associated KK relations can also be fleshed out  \cite{Jia:2010nz}. The BCFW shift of adjacent gluons on tree amplitudes is given by equation \eqref{eq:adscale} for renormalisable couplings of scalar and fermionic adjoint matter \cite{Cheung:2008dn}. Hence the entire argument just constructed goes through for this class of theories, assuming color-kinematic duality holds: the gravity amplitudes constructed by double copy scale simply as the scaling of the double copy of the gauge theory components as in \eqref{arkanigrav}. Actually, also shifts of other particles than gluons can be straightforwardly be `squared' this way.  

\subsection*{Improved scaling for permutation sums} 

One simple application of the results of this section to tree level amplitudes follows from considering BCFW shifts of two gluons color-adjacent either to a permutation or a cyclic sum over external legs. Inspired by earlier work in \cite{Badger:2008rn} it was further fleshed out in \cite{Boels:2011tp, Boels:2011mn} that these shifts show improved shift behavior. A shift of a pair of color-non-adjacent particles is a particular example of this class of shifts. The improved shift behavior has since been proven in \cite{Du:2011se} and \cite{Du:2011fc} using the BCJ relations for scattering amplitudes. Since the BCJ relations are a consequence of color-kinematic duality, it is not surprising that the same conclusion follows directly from the results obtained above on shifts of numerators. 

The point is that the particles within the permutation or cyclic sum have to couple to the hard line directly: a permutation sum on a current with more than one on-shell particle vanish by the photon decoupling relation. If the amplitude is now expressed in terms of the color ordered version of the BCJ representation, it is clear that every additional particle on the hard line will lead to a $\frac{1}{z}$ suppression from the `propagator' type terms since only trivalent graphs are involved. By the results just derived the numerators scale as the Yang-Mills tree amplitude up to terms which do not contribute to the Yang-Mills amplitude. Hence the result for suppression conjectured in \cite{Boels:2011tp, Boels:2011mn} follow immediately. In a real sense, this is the color-kinematic counterpart of the proof in  \cite{Du:2011se} and \cite{Du:2011fc}. 

\subsection*{Further comments}

It should be obvious that the results of supersymmetric generalization of BCFW shifts considered in \cite{Brandhuber:2008pf} (see also \cite{ArkaniHamed:2008gz}) for gravity amplitudes also simply follow from the above argument. Finally, nothing in this section depends crucially on selecting the $\textrm{D}^3\textrm{M}$ basis: any basis will do. Actually, from a formal point of view it is somewhat more natural not to solve the Jacobi's and simply treat Jacobi's and equation \eqref{eq:BCJtree} as a massive set of linear equations which may be inverted using generalized inverses. A similar comment applies to the basis of trivalent scalar theories used to study the scaling of the generalized inverse. Although above a choice of $(n-3)!$ different numerators not in the kernel was made, for the argument to go through it is not needed to really specify this number: one could take an over-complete basis. The expansion coefficients $\alpha$ are then not unique, but this can with a bit of tedious argument through generalized inverses be shown not to influence the result.


\section{BCFW shifts of gravity integrands from gauge theory}\label{sec:integrands}

Given the above results on BCFW scaling of Einstein gravity tree amplitudes through color-kinematic duality the next step is to extend the analysis to the integrand level. It will be argued that essentially all the steps of the tree level derivation go through, up to several new subtleties/assumptions that will be identified and addressed when needed. They are given by
\begin{enumerate}[(I)]\label{assumplist}
\item Color-dual numerators exist (in particular any consistency conditions on the Yang-Mills integrands are fulfilled).
\item The trivalent scalar theories span the set of integrands which satisfy the consistency conditions.
\item In the chosen color-basis the coefficients of the integrand for trivalent scalar theories scale the same or worse (higher powers in $z$) compared to the corresponding coefficient for the Yang-Mills integrand.
\item Generalized gauge transformations involved do not influence the double copy relations.
\end{enumerate}

\subsection{BCFW shifts of kinematic numerators using generalized inverses}

The integrand of an $l$-loop  $n$-point Yang Mills amplitude can be written in a cubic representation, just as the tree level amplitude can through equation \eqref{eq:BCJintegrand}, reproduced here for convenience: 
\begin{equation}\nonumber
\mathcal{A}^l_n= g_{ym}^{n-2+2l}  \int \prod_{j=1}^l d^DL_j  \sum_{\Gamma_i} \frac{1}{S_i} \frac{n_i c_i}{s_i}
\end{equation}
As before, the color factors $c_i$ in this representation obey a set of Jacobi relations and following assumption (I) (i.e.\ hypothesis \ref{susp:1}) it will be assumed that a set of kinematic numerators $n_i$ can be found that obeys corresponding Jacobi relations. 

As before, these relations may be solved in terms of $h$ different basis color factors and numerators,  denoted by $\breve{c}$ and $\breve{n}$. At one loop for instance, the $\textrm{D}^3\textrm{M}$ basis of equation \eqref{KKbasis1loop} is an explicit example of such a solution with dimension $h=(n-1)!/2$. Inserting the basis into the cubic representation it can be rewritten analogous to the tree level case as
\begin{equation}\label{bcjsatisintegrand}
\mathcal{A}^l_n \sim \int \prod_{j=1}^L d^DL_j \sum_{\Gamma_i}\frac{1}{S_i} \frac{n_i c_i}{s_i} \rightarrow \breve{c}^k \int \prod_{j=1}^l dL_j  F_{km} \breve{n}^m
\end{equation}
where $k$ and $m$ run from 1 to $h$ and $F$ is a matrix whose entries are sums over products of scalar propagators. The matrix $F$ is not unique but depends column by column on the definition of loop momentum. These definitions have to be taken into account carefully when summing up the scalar propagators. The left hand side can also be expressed in the chosen color basis
\begin{equation}
\mathcal{A}^l_n  =  \breve{c}^k \int \prod_{j=1}^l d^DL_j\, I^{YM}_k 
\end{equation}
It can be useful to think of the integrand as defined at least in principle through Feynman graphs. A new feature at loop level is that one \emph{cannot} extract the equation
\begin{equation}
 I^{YM}_j = F_{jm} \breve{n}^m \qquad \textrm{(does not hold)}
\end{equation}
as a consequence of equation \eqref{eq:BCJintegrand}. The main difference to tree level is that while the numerators live in the space of vectors of functions of external and loop momenta,
\begin{equation}
n \sim \left(\begin{array}{c} f_1(p_i, L_i) \\ f_2(p_i, L_i)\\ \ldots \end{array} \right)
\end{equation}
 the integrands live in the space of vectors of functions of external and loop momenta, identified up to terms which integrate to zero. That is,
\begin{equation}
I \sim \left(\begin{array}{c} f_1(p_i, L_i) \\ f_2(p_i, L_i)\\ \ldots \end{array} \right)
\end{equation}
where two functions $f$ and $g$ are equivalent, $f\sim g$ if 
\begin{equation}
 \int \prod_{j=1}^l d^DL_j f =  \int \prod_{j=1}^l d^DL_j g
\end{equation}
The space of integrands is therefore in a real sense smaller than the space of numerators. As a linear map the matrix $F$ therefore can have a non-trivial kernel.

Some of the ambiguity has to do with routing the loop momenta on the left and right hand side. For convenience it is natural to involve the same choice on the two sides as will be enforced below. For a BCFW shift this implies on both sides there a hard line is chosen as the minimal length path between the shifted legs. In general the equation to solve is
\begin{equation}
 I^{YM}_j \sim F_{jm} \breve{n}^m
\end{equation}
or equivalently in the space on which the numerators live
\begin{equation}
 I^{YM}_j+I_j^{vanish} = F_{jm} \breve{n}^m
\end{equation}
For a solution  to this equation to exist there could be consistency conditions if there are left null eigenvectors of $F n$. Note that `null' in this sentence is up to terms which vanish after integration. Actually, at one loop we strongly suspect the relations found in \cite{Boels:2011tp} are the full set of consistency conditions for local numerators. At higher loops similar relations may exist as argued in \cite{Boels:2011tp, Boels:2011mn} on the basis of the improved behavior under non-adjacent BCFW shifts.

In the linear map approach the assumption (I) that color-dual numerators exist is simply the assumption that all consistency conditions are satisfied. This is the analogue of equation \eqref{eq:conditiononamps} at tree level. If this holds then by equation \eqref{eq:geninvmaps} the numerators at loop level are given by
\begin{equation}
\breve{n} = F^+ ( I^{YM}) + \textrm{ker} F
\end{equation}
In the tree level case the terms in the kernel of $F$ were generalized gauge transformations which can be shown not to affect the squaring relation. As discussed above in section \ref{sec:rev} the same does not hold for integrand numerators. This is a problem inherent to the color kinematic duality. It will be assumed this issue will not affect the outcome of the double copy relation for the integrand in a meaningful way in the following analysis. 

By assumption (II) the non-trivial eigenvectors of $F$ are spanned by trivalent scalar theories. This amounts to the assumption that the space of solutions to the consistency conditions at loop level for color-kinematic duality is spanned by this class of theories. At one loop it is certainly true that both trivalent scalar theories as well as Yang-Mills theories obey the relation found in  \cite{Boels:2011tp}. Moreover, both show a factor of $\frac{1}{z}$ improvement for non-adjacent over adjacent shifts.  In equations this assumption states that the left-side kernel of the map $F$ 
\begin{equation}
\textrm{left ker} F = \{m_j | m_i F^{ij} \breve{n}_j = 0\}
\end{equation}
is the same set for numerators from the trivalent scalar theories as well as Yang-Mills theory\footnote{Some evidence for this will be provided later by an estimate of the kernel dimensions in both cases from BCFW shifts.}.

With this assumption the scaling of the generalized inverse $F^+$ up to gauge transformations can be studied as before by expanding the Yang-Mills numerators as
\begin{equation}\label{eq:expintegrandnuminalpha}
\breve{n}_i=\alpha_{\bar{K}} \breve{n}^{\bar{K}}_i
\end{equation}
where 
\begin{equation}
F_{ji} \breve{n}^{\bar{K}}_i \neq 0
\end{equation}
holds and the numerators are taken to be linearly independent. Here unbarred indices run from $1$ to the number of color-ordered integrands independent under (suitably generalized loop level) KK relations, $h$, and barred indices from $1$ to the number of independent color-ordered integrands under the consistency conditions (the analogs of the BCJ relation at tree level) at the integrand level.  This gives an expansion for the integrand of Yang-Mills theory as 
\begin{equation}\label{eq:expintinalpha}
(I+I^{\text{vanish}})_i =\alpha_{\bar{J}}\Theta^{\bar{J}}_i
\end{equation}
where $\Theta^{\bar{J}}_i$ is the integrand of the $\overline{J}$th trivalent scalar field theory. As before, the action of any generalized inverse $F^+$ on $\Theta^{\bar{J}}_i$ can be inferred from the general solution
\begin{equation}
\breve{n}^{K} = F^+ \Theta^{\bar{K}} + \textrm{ker} F
\end{equation}
In trivalent scalar theories the left hand side scales as $z^0$ up to a generalized gauge transformation. The right hand side can be more complicated. In a typical choice of color basis the shift of two legs the elements of the integrand vector are linear combinations of color ordered integrands. At one loop for instance in the $\textrm{D}^3\textrm{M}$ basis a shift of two legs will be either adjacent or non-adjacent. Hence for the class of trivalent scalar theories
\begin{equation}
F^+ \Theta \sim \left(\begin{array}{cc} \sim z^0 & \sim z^1 \\ \sim z^0 & \sim  z^1 \end{array} \right)  \left(\begin{array}{c} \sim z^0  \\ \sim   z^{-1} \end{array} \right) 
\end{equation}
up to a generalized gauge transformation at one loop. Note the improvement in scaling of the integrand for non-adjacent shifts leads to a less well behaved generalized inverse $F^+$. Now for the Yang-Mills integrand at one loop one immediately obtains 
\begin{equation}
n^{(1)}_{ym} = F^+ I^{(1)}_{ym}  \sim  \varepsilon(\hat{1})_\mu\varepsilon(\hat{n})_\nu \Big(z \eta^{\mu\nu}f(1/z)+z^0B^{\mu\nu}\Big)+\mathcal{O}(1/z) + \textrm{kernel}
\end{equation}
from the results for adjacent and non-adjacent shifts for loop level integrands. These results are obtained up to contribution which vanish after integration. If these would scale worse than the behavior indicated in the equation, then these would satisfy the consistency conditions themselves. Hence these are generalized gauge transformations.

At higher loops the same argument goes through under an additional assumption (assumption (III)): if one of the coefficients in the integrand vector scales shows improved BCFW scaling behavior, then the corresponding gauge theory coefficients must do so as well. Generically these coefficients are expected to either involve adjacent or non-adjacent shifts for which the shift behavior is known. Also, from the results of \cite{Naculich:2011ep, Edison:2011ta, Edison:2012fn} there is a (conjectured) basis for $4,5,6$ point integrands which is simply a subset of the full set of color-ordered integrands. This is enough to imply the assumption. 

Up to the assumptions  the scaling of the numerators up to generalized gauge transformations is given by
\begin{equation}\label{numeratorsanyloop}
 \lim_{z\rightarrow\infty} n^k \sim  \varepsilon(\hat{1})_\mu\varepsilon(\hat{n})_\nu \Big(z \eta^{\mu\nu}f(1/z)+z^0B^{\mu\nu}\Big)^k+\mathcal{O}(1/z) + \textrm{kernel}
\end{equation}

\subsection{BCFW shifts of gravity integrands constructed by double copy} 
The double copy relation at loop level can be rewritten in terms of the linear map approach from equation \eqref{eq:BCJloopssquaring}. This reads
\begin{equation}\label{eq:BCJloopssquaringmaps}
 M_n^l  \sim \int \prod_{j=1}^l d^DL_j \sum_{\Gamma_i} \frac{1}{S_i} \frac{n_i \tilde{n}_i}{s_i} \sim \int \prod_{j=1}^l d^DL_j\; \tilde{\breve{n}}^i F_{ij} \breve{n}^j
\end{equation}
where $\tilde{\breve{n}}$ and $\breve{n}$ are sets of kinematic numerators from two copies of gauge theory. Knowing the large-$z$ behavior of the numerators at any loop order \eqref{numeratorsanyloop} and of $F$ it follows immediately that under a shift of particles $1$ and $n$ the large-$z$ scaling of the $n$-graviton gravity integrand $\mathcal{I}_n$ is given by a double copy of \eqref{numeratorsanyloop}
\begin{equation}\begin{split}\label{gravintegrandscale}
 \lim_{z\rightarrow \infty}\mathcal{I}_n \sim
\varepsilon(\hat{1})_\mu \varepsilon(\hat{n})_\nu \tilde{\varepsilon}(\hat{1})_{\tilde{\mu}}\tilde{\varepsilon}(\hat{n})_{\tilde{\nu}} \Big(z^2 \eta^{\mu\nu}\tilde{\eta}^{\tilde{\mu}\tilde{\nu}}f(1/z)+z(\eta^{\mu\nu}\tilde{B}^{\tilde{\mu}\tilde{\nu}}+B^{\mu\nu}\tilde{\eta}^{\tilde{\mu}\tilde{\nu}})+\\z^0(B^{\mu\nu}\tilde{B}^{\tilde{\mu}\tilde{\nu}})+\frac{1}{z}(B^{\mu\nu}\tilde{B}^{\tilde{\mu}\tilde{\nu}})\Big)+\mathcal{O}(1/z^2)
\end{split}
\end{equation}
up to the assumptions above. In particular, it is equivalent to the tree level result \eqref{arkanigrav}. Let us stress again that this result is under the assumption (IV), i.e.\ the generalized gauge transformation encountered along the way do not interfere with the double copy formula.

Note this result displays even more dramatic cancellations in the powers of $z$ than at tree level: at $l$ loop for $n$ gravitons the large-$z$ scaling should na\"ively be given by $\sim z^{n-2+2l}$ but as one can observe from the above formula the largest power in $z$ is quadratic and moreover independent of the loop order. As in the tree level case this makes it explicit that the driving force behind these cancellations is the assumed existence of color-kinematic duality.

\subsubsection*{Comments}
The large-$z$ scaling result above \eqref{gravintegrandscale} for the $\mathcal{N}=0$ integrand remains valid if scalars or fermions in the adjoint are included. This follows from the argument for adding matter at tree level in section $3$. The needed results for scaling of Yang-Mills integrands can be found in  \cite{Boels:2010nw, Boels:2011mn}. Similarly, the comments in section \ref{sec:amplitudes} on supersymmetric shifts and some minor variations also apply to the arguments presented here. 

From the scaling of the inverse of $F$ in the integrand case one can estimate the number of independent integrands under the analog of the BCJ relation at loop level. This follows as the dimension of the kernel of $F$ does not change in the limit: this is basically equivalent to the statement that $F^+(z) \sim z^0$ up to a generalized gauge transformation. In the large $z$ limit in a scalar theory it is straightforward to estimate the number of independent integrands for $n$ external legs under the BCJ relations is the equal to the number of independent integrands under the KK relation for $n-1$ particles. This follows from the fact that the leading diagram in the large $z$ limit is a Feynman graph coupling to a one-leg off-shell current which involves $n-2$ on-shell particles. For these the analog of the KK relation holds, while no additional BCJ relation are expected for this off-shell object. Note in particular that both in the trivalent scalar theories as well as in Yang-Mills theory these results are basically the same. See also section \ref{sec:feynman} for the gauge theory analysis. This provides a cross-check of the expectation that the consistency conditions for the scalar theories are the same as those for the Yang-Mills theory. 



\section{Estimates of UV divergences assuming color-kinematic duality}\label{sec:uv}

The main motivation for the study of cancellations in gravity theories is the UV behavior of the theory. Here we will only study the overall degree of divergence of graphs, ignoring sub and overlapping divergences. In other words, loosely speaking we study the limit
\begin{equation}
L_i \rightarrow \infty \qquad \forall i
\end{equation}
The techniques based on linear maps employed above for the study of BCFW shifts can also be applied to the study of UV divergences through powercounting. It will be argued that color-kinematic duality leads directly to improved powercounting behavior of gravity integrands, subject to the same assumptions as listed at the beginning of  the previous section adapted with respect to the large loop momentum. It should be noted a natural expectation present in much of the literature on divergences is that if a certain divergence is absent, a form of the integrand exists which makes this absence manifest under this scaling. If powercounting on the other hand yields a possible divergence, this could still vanish after integration and summation of all the terms.

Note that it is known that the study of large BCFW shifts is related to the ultraviolet degrees of divergence, even beyond the fact that they are both based on powercounting. The canonical example of this is the good behavior of gravity tree amplitudes under BCFW shifts which is directly responsible for the absence of scalar triangle integrals at one loop in $\mathcal{N}=8$ supergravity \cite{ArkaniHamed:2008gz} (see also \cite{BjerrumBohr:2008ji}). Hence by the results obtained above for tree amplitudes the absence of triangles at one loop for any number of points can by the results of this article be understood in terms of symmetry: it is a consequence of color-kinematic duality. Through unitarity, this argument factors through to the integrand at any loop order: there can be no triangle graphs, or at least none that can be isolated by a unitarity cut. In particular there can be no triangle divergences in either maximally supersymmetric Yang-Mills or gravity theory. 

For the purposes of this section one can re-write equation \eqref{eq:BCJloopssquaring} as
\begin{equation}\label{eq:asymsquare}
 M_n^l  \sim \int \prod_{j=1}^l d^DL_j \sum_{\Gamma_i} \frac{1}{S_i} \frac{\breve{n}_i \tilde{\breve{n}}_i}{s_i} \sim \int \prod_{j=1}^l d^DL_j\; \tilde{\breve{n}}^i I_i
\end{equation}
with $I_i$ one of the copies of the gauge theory integrand, while the other copy yields the numerators $\breve{n}$.

Before we study the UV behavior of gravity theories, let us briefly recapitulate the UV behavior of (supersymmetric) Yang-Mills theories as these is the input in the double copy construction.
At one loop it is well known that maximally supersymmetric Yang-Mills theory diverges in $D=8$. More generically, the usual decomposition of integrands of gauge theories with massless matter in terms of massless box, triangle and bubble integrals shows clearly the UV divergences in these theories at the bubbles\footnote{Interestingly, this relates sums over bubble coefficients to tree level amplitudes, see \cite{ArkaniHamed:2008gz} and especially \cite{Huang:2012aq}.}.

At higher ($>1$) loops the critical dimension where maximally supersymmetric Yang-Mills diverges logarithmically is \cite{Galperin:1984bu},  \cite{Bern:1998ug}
\begin{equation}
D_c=\frac{6}{L} +4
\end{equation}
This has been explicitly shown to be saturated up to including five loops \cite{Bern:2012uc} for general color and to six loops in the planar sector \cite{Bern:2012di}. In particular in four dimensions this theory is UV finite. This translates to a power count of the integrand of amplitudes in maximally supersymmetric Yang-Mills theory of
\begin{equation}\label{eq:uvdivergencemaxYM}
\int d^{D l} L \frac{(L^{2})^{l-2}}{(L^2)^{3l+1}}
\end{equation}
at loop order $l$. The number of propagator powers is that of a four point amplitude with the maximal number of loop propagators.

In pure Yang-Mills the situation is different. Here na\"ive powercounting of Feynman graphs gives a logarithmic divergence in two dimensions. This divergence is absent in dimensional regularization. This can be seen from the background field method: the functional form of the two-point correlator of two (background) gauge fields which contains the two dimensional divergence has to be proportional to
\begin{equation}
\langle A_{\mu} A_{\nu} \rangle \sim g_{\mu\nu} p^2 + p_{\mu} p_{\nu}
\end{equation}
by the background field gauge invariance. The two-dimensional divergence would arise with the first term, while the second cannot arise with a two-dimensional divergence. Hence by gauge invariance the two dimensional divergences must vanish sum and consequently generic Yang-Mills theories are logarithmically divergent in four dimensions.

The main observation in this section is that the analysis of large BCFW shifts using color-kinematic duality of the previous section can be adapted to the limit of large loop momenta. More precisely, the inner product of all loop momenta with external momenta is taken to zero. As this is a well-defined limit, it should have a reflection for the kinematic numerators, up to the subtlety of the generalized gauge transformations. This limit will allow us to provide an estimate the overall degree of divergence of the corresponding gravity theory.

As already mentioned, the assumptions at the core to powercounting UV analysis at loop level are basically those at the beginning of the previous section \ref{assumplist} with large-$z$ behavior replaced by the large-loop momentum limit.
The main difference to BCFW shifts is that where in that case also tree level propagators are involved, here the tree propagators do not contribute to UV divergences as they do not involve loop momenta.  With the above assumption the limit behavior of the linear map $F^+$ can be studied in the class of trivalent scalar theories. Since the numerators of these theories do not depend on loop momenta, this determines the scaling of $F^+$ up to generalized gauge transformation from 
\begin{equation}
\breve{n}^{\bar{K}} = F^+ ( I^{\textrm{trivalent}, \bar{K}}) + \textrm{ker} F
\end{equation}
for the $\bar{K}$-th trivalent scalar theory. A subtlety arises here as the scaling of $F^+$ is now determined from the scaling of the different members of the minimal-under-KK-relations integrand basis $I$. As argued before, one version of this basis is likely to be a subset of all color-ordered integrands. One would now expect the multi-trace terms at least to have better powercounting than the planar ones. In pure Yang-Mills this is related to renormalizability: the UV divergences must be such that they can be reabsorbed into a counterterm proportional to the Yang-Mills action. Hence, these divergences must be proportional to tree amplitudes and in particular be single trace. 

If all the integrands in the trivalent scalar theory basis  the integrands scale as $\sim (L^2)^{-\delta}$ for some integer $\delta$. Then $F^+ \sim (L^2)^{\delta}$. However, if the integrand vector scales partly as, say, $(L^2)^{-\delta_1}$ and partly as $(L^2)^{-\delta_2}$, then the generalized inverse scales $F^+ \sim (L^2)^{\delta_1}$ or  $F^+ \sim (L^2)^{\delta_2}$, depending on the column number of $F^+$. In schematic equations,
\begin{equation}\label{eq:differentlyscalingnums}
\textrm{if} \qquad I \sim \left(\begin{array}{c}(L^2)^{-\delta_1} \\(L^2)^{-\delta_2} \end{array} \right) \qquad \textrm{then} \qquad F^+ \sim \left(\begin{array}{cc}(L^2)^{\delta_1}  & (L^2)^{\delta_2} \end{array} \right)
\end{equation} 
The scaling of $F^+$ is as ever up to generalized gauge transformations. This behavior then determines the powercounting behavior of the numerators of the gauge theory by expanding them in the trivalent scalar theory basis.

The next step is to decide which trivalent graphs to include in equation \eqref{eq:BCJintegrand} or put differently, which graphs come with non-zero numerators. This is less innocuous than it sounds. For maximally supersymmetric Yang-Mills theory for instance it is extremely natural to exclude all triangle graphs. Apart from yielding a restricted set of graphs this also simplifies the Jacobi identities. In particular, this will admit a solution of the numerator Jacobis which is smaller than the one obtained for color Jacobis. In other words, there is a rectangular matrix such that
\begin{equation}\label{numsubset}
\breve{n}_j = N_{jM} \bar{n}^M
\end{equation}
Here $j$ ranges over all elements of the set of solutions to the color Jacobi relations. Note that in known examples, see e.g. \cite{Bern:2010ue}, all Jacobi numerators can be expressed in terms of just one or two `master' graphs. This limits the range of the index $M$, $\#M$, to be $<< \#j$. It would be very interesting to understand this from a group theory point of view. Many formulas given above can be streamlined in terms of the $\bar{n}$ basis, but this will mostly not be essential for this section.

\subsubsection*{One loop}
At one loop, excluding triangle graphs gives for maximally supersymmetric Yang-Mills theory
\begin{equation}
F^+ \sim (L^2)^4 \qquad \mathcal{N}=4, 1 \textrm{ loop}
\end{equation}
from the trivalent scalar theories, as all the coefficients in the integrand vector in the $\textrm{D}^3\textrm{M}$ basis contain at least a box. This gives
\begin{equation}
n_{ym} \sim (L^2)^0 \qquad \mathcal{N}=4, 1 \textrm{ loop}
\end{equation}
A very similar analysis for less-supersymmetric Yang-Mills yields
\begin{equation}
F^+ \sim (L^2)^2 \qquad \mathcal{N}<4, 1 \textrm{ loop}
\end{equation}
which implies
\begin{equation}
n_{ym} \sim (L^2)^0 \qquad \mathcal{N}<4, 1 \textrm{ loop}
\end{equation}
Combined with the double copy formula \eqref{eq:asymsquare} this immediately gives the result that the double copy formula will generically scale as the best-behaved gauge theory copy. In particular, it implies the no-triangle property of $\mathcal{N}=8$ supergravity as a consequence of that of $\mathcal{N}=4$ super Yang-Mills through the duality. Moreover, it correctly gives a logarithmic divergence in $\mathcal{N}=0$ supergravity in four dimensions as well as the correct critical dimension (D=8) for $\mathcal{N}=4$ supergravity at one loop. In other words, at this loop order all known critical dimensions in the literature are reproduced, under the assumptions listed above. Of course, the analysis here does not include a check if the critical dimension is saturated. 

These results for scaling behavior are, as always, up to generalized gauge transformations. Moreover, if there is a vanishing integrand involved in the connection to the numerators, potential worse scaling behavior of this integrand can as in the BCFW shift case be absorbed into a generalized gauge transformation.

\subsubsection*{Higher loops}
\begin{figure}[b]
\centering
\includegraphics[scale=0.9]{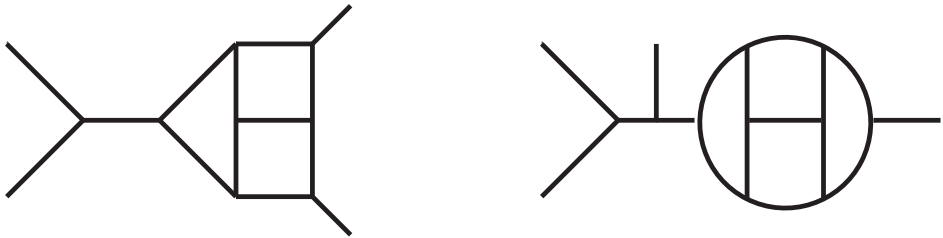}
\caption{\label{fig:contact}Contact graphs appearing at three loops (left) and at four loops (right).}
\end{figure}

At higher loops the graph topologies appearing in  the duality have in general a large effect on the generalized inverse of the linear map $F$, even apart from the issue sketched in \eqref{eq:differentlyscalingnums} of differently scaling trivalent scalar theory integrands. For this consider $\mathcal{N}=4$ at four points. If one restricts to trivalent graphs with the maximal number of loop propagators, 
\begin{equation}
F^+ \sim (L^2)^{3l+1} \qquad \mathcal{N}=4, \textrm{max \# loop propagators}
\end{equation}
and consequently
\begin{equation}
n_{ym} \sim (L^2)^{l-2} ,  \qquad \mathcal{N}=4, \textrm{max \# loop propagators}
\end{equation}
from $l=2$ loops onwards. After squaring, this leads to a three loop estimate of the critical dimension of $\mathcal{N}=8$ which is too divergent compared to the known answer. Luckily, the answer can be seen from the known color-dual form of the three loop integrand in \cite{Bern:2010ue}: there are graphs such as on the left hand side of figure \ref{fig:contact} which have one loop propagator less than the maximal number. Including these in the trivalent graphs and assuming all integrands scale homogeneously in the large loop momentum limit gives 
\begin{equation}
F^+ \sim (L^2)^{3l} \qquad \mathcal{N}=4, \textrm{1  loop propagator less}
\end{equation}
and consequently
\begin{equation}
n_{ym} \sim (L^2)^{l-3} ,  \qquad \mathcal{N}=4, \textrm{1  loop propagator less}
\end{equation}

Similarly, at four loop graphs of the type on the right hand side of figure \ref{fig:contact} appear in the explicit result for the four point integrand \cite{Bern:2012uf} . This kills of an additional power of $L^2$, leading to
\begin{equation}
F^+ \sim (L^2)^{3l-1} \qquad \mathcal{N}=4, \textrm{2 loop propagators less}
\end{equation}
and consequently
\begin{equation}
n_{ym} \sim (L^2)^{l-4} ,  \qquad \mathcal{N}=4, \textrm{2 loop propagators less}
\end{equation}
This gives an estimate of the critical dimension of $D_c = 5.5$ of $\mathcal{N}=8$ supergravity at four loops. Note that up to and including four loops it is highly plausible by the above argument that the numerators in the higher point integrands obey the same scaling relations and that, consequently, the critical dimension for these integrands is the same.  

At five loops a problem begins to set in: if $\mathcal{N}=8$ has to have the same critical dimension as $\mathcal{N}=4$\footnote{It should be stressed here it could still be true that the powercounting gives too pessimistic estimates.}, then for four points a graph with $3$ tree propagators is necessary. Restricting to graphs without internal triangles, the only possibility for this is a tadpole. We have explicitly verified that four possible five-loop tadpole graphs made only out of box and pentagon type internal graphs exist using DiaGen \cite{DiaGen}. If these tadpole graphs appears in the sum of equation \eqref{eq:BCJintegrand}, then the critical dimension by the power count estimate described here is the same as $\mathcal{N}=4$ super Yang-Mills at five loops. If these tadpole graphs do not appear, this leads directly to an expected seven loop divergence in $\mathcal{N}=8$ supergravity in four dimensions. If the five-loop tadpole appears and no further improvements occur then the above reasoning leads to the possibility that  $\mathcal{N}=8$ supergravity diverges in four dimensions at $8$ loops. 

\begin{figure}[h!]
\centering
\includegraphics[width = 3.6in, height = 1.51in, scale=0.5]{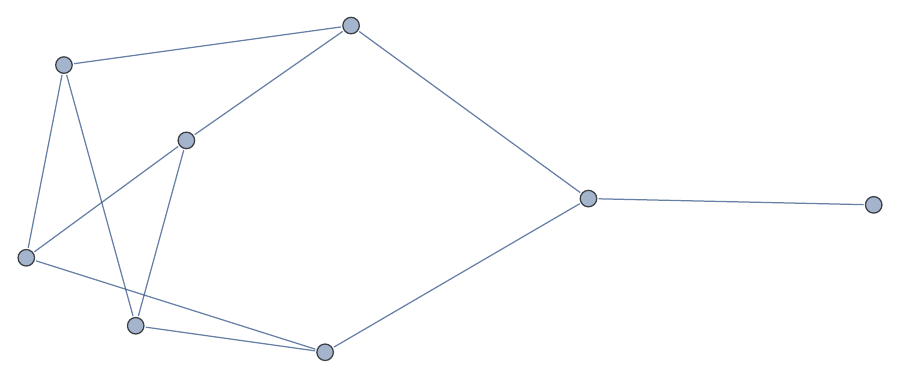}
\caption{\label{fig:4looptadpole} Possible four loop tadpole graph made of boxes.}
\end{figure}

There are several ways to improve overall UV degree of divergence behavior. One is to include the possible four loop tadpole graph which consists of boxes only, see figure \ref{fig:4looptadpole}. If this graph appears, then the above reasoning estimates that $\mathcal{N}=8$ supergravity diverges in four dimensions at $10$ loops. If one relaxes the constraint to graphs without internal triangles, there are no problems in finding trivalent graphs to produce more cancellations. For $\mathcal{N}=8$ to have the same power count as $\mathcal{N}=4$ SYM to all loop orders, one should include trivalent graphs which have the same UV power count as the $\mathcal{N}=4$ answer. Of course, the numerator factors here should be such that no triangle subgraph can be isolated by unitarity cuts in the gauge theory integrand.

\subsubsection*{Powercounting in half-maximal supergravity} 
The above methods can also be applied to study gravity integrands in gravitational theories with less than $\mathcal{N}=8$ supersymmetries. One well-studied theory is half-maximal supergravity. It was recently shown to diverge in $D=8$ at one loop and $D=6$ at two loops. Moreover, it was shown to be finite in $D=4$ at three loops \cite{Bern:2012gh, Bern:2012cd, Tourkine:2012ip}. The gravity integrand of $\mathcal{N}=4$ can be obtained via the double copy construction as the product of $\mathcal{N}=0$ and $\mathcal{N}=4$ Yang-Mills in two different ways,
\begin{equation}\label{gravhalfsusy1}
M_n^l \sim \int \prod_{j=1}^l d^DL_j I^{\mathcal{N}=4}n^{\mathcal{N}=0}\sim \int \prod_{j=1}^l d^DL_j I^{\mathcal{N}=4} F^+_{\mathcal{N}=0}I_{\mathcal{N}=0}
\end{equation}
or
\begin{equation}\label{gravhalfsusy2}
M_n^l \sim \int \prod_{j=1}^l d^DL_j I^{\mathcal{N}=4}n^{\mathcal{N}=0}\sim \int \prod_{j=1}^l d^DL_j I^{\mathcal{N}=0} N F^+_{\mathcal{N}=4}I_{\mathcal{N}=4}
\end{equation}
where $N$ is the matrix out of equation \eqref{numsubset} which connect $\mathcal{N}=4$ and $\mathcal{N}=0$ kinematic numerators. Note that the first formula gives a critical dimension of $D=8$ by the result listed above, while the second gives $D_c=4$. Moreover, since the $\mathcal{N}=4$ SYM numerators are known to scale as $(L^2)^0$ at least to four loops, the latter formula estimates $\mathcal{N}=4$ supergravity to be logarithmically divergent up to four loops. However, more cancellations can be hidden in the sums, just as they are at one loop. 

From the first formula it follows that the scaling of the pure Yang-Mills numerators sets the critical dimension of $\mathcal{N}=4$ supergravity compared to $\mathcal{N}=4$ super Yang-Mills. Naive expectation would be that these numerators scale as $\sim (L^2)^0$ if all integrands in the color basis scale the same which would give $\mathcal{N}=4$ supergravity the same UV divergence as $\mathcal{N}=8$, while it is known that the $2$ and $3$ loop critical dimensions are $6$ and $\sim 4$ respectively. At two loops, this translates into one power of $L^2$. This can be argued to arise as follows: two loops is the first instance multi-trace terms appear which cannot related to the single trace term. In Yang-Mills these are expected to scale better in the large loop momentum limit by renormalizability of the theory in $4$ dimensions. In the basis of trivalent scalar theories, the leading contribution to the triple trace terms contains three loop propagators more: there may not be internal color triangle or bubble terms as they are proportional to either the color $\delta^{ab}$ or $f^{abc}$. Hence inverting this and taking into account the better scaling of Yang-Mills easily gives additional powers of $L^2$. In fact, from the $\mathcal{N}=4$ supergravity result it is easy to infer that the triple trace integrand in pure Yang-Mills at two loops should scale two powers of $L^2$ better than the planar counterpart. 

\subsubsection*{A sketch of renormalizability in Yang-Mills}  
Gravitational theories with less than $\mathcal{N}=4$ supersymmetry are expected to diverge in four dimensions. Moreover, generic Yang-Mills theory is logarithmically divergent but renormalizable in four dimensions. In this short subsection we sketch an outline of how renormalizability could work in $\mathcal{N}=0$. 

\begin{figure}[b]
\centering
\includegraphics[scale=0.8]{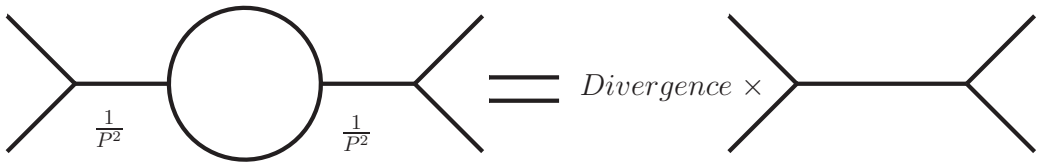}
\caption{\label{fig:divergence} The bubble graph divergence at one loop}
\end{figure} 

Since at one loop the numerators scale as $(L^2)^0$, the only UV divergent graph in the color-kinematic dual integrand in four dimensions is a bubble presented in figure \ref{fig:divergence}. Based on dimensional analysis, the divergence can only be proportional to $p^2$, where $p$ is the momentum flowing into the bubble. This cancels off one of the two $\frac{1}{p^2}$ propagators. Hence the UV divergence at one loop must by the results obtained here be proportional to a sum over trivalent graphs which looks like the color-dual representation of the tree level amplitude. This is the hallmark of renormalizability. Note that one can also conclude that for the quadratic Casimir for the kinematic algebra scales as $C_2 \sim p^2 + \mathcal{O} ((L^2)^{-1})$.

\begin{figure}[h!]
\centering
\includegraphics[scale=0.5]{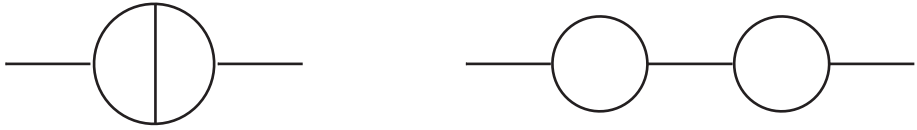}
\caption{\label{fig:twoloops} Divergent trivalent graphs at two loops.}
\end{figure}

The same reasoning immediately extents to concatenated bubble graphs at higher loop orders. There can be more graphs which contribute at higher loops, see figure (\ref{fig:twoloops}). Moreover, it is known the $\beta$ function of Yang-Mills theories contains non-planar corrections \cite{vanRitbergen:1997va}, see also appendix $B$ in \cite{Boels:2012ew} for a discussion. These graphs should also appear, with the value of the corresponding higher Casimirs for the kinematical algebra fixed, again, by dimensional analysis. The sum over all divergences should, at least in color space, reduce to a divergence times the corresponding tree amplitude. It would be interesting to pursue this further.

Note that squaring the one-loop Yang-Mills result gave a logarithmic divergence in $\mathcal{N}=0$ supergravity in four dimensions at one loop, as expected. Interestingly, taking the simple square of the Yang-Mills numerator scaling and analyzing the resulting graphs as above gives instead of a tree level-type graph with a $\frac{1}{p^2}$ propagator a contact-type graph. This is very much like what one would expect from a local counterterm different than the Einstein-Hilbert action. It would be interesting to see what the expression for this graph gives. 

From the $2$ loop $\sim (L^2)$ behavior of the pure Yang-Mills numerators extracted from the $\mathcal{N}=4$ supergravity result one would naively expect a quadratic divergence at two loops in $\mathcal{N}=0$ supergravity, driven by the triple trace parts of the Yang-Mills theory. However, since these arise in the non-planar sector of the theory, there is an argument to be made that the actual divergence is milder. It would be interesting to study this further.


\section{Towards an off-shell understanding of numerator scaling}\label{sec:feynman}
In this section it will be argued that the large-$z$ behavior of the kinematic numerators at tree and integrand level can actually be seen from Feynman diagrams directly using standard power counting. The key point will be to use color Jacobi relations to relate different contributions so that cancellations are achieved in the sum over Feynman graphs and the large-$z$ scaling of the kinematic numerators found in the previous sections will be -- at least conceptually -- reproduced.

\subsection{Comparison to BCFW shifts directly via Feynman graphs}
To obtain the large-$z$ shift of kinematic numerators from Feynman diagrams directly Feynman rules in the Feynman-'t Hooft gauge will be used and the gluon propagator will be put in the lightcone AHK gauge $q\cdot\mathcal{A}=0$ (see figure \ref{fig:powercountingfeynmanrules}). An advantage of using this gauge is that if all the unshifted legs are left off-shell the scaling obtained holds at the level of the integrand to all loop orders. For brevity only the leading large-$z$ contribution will be considered in the following but the subleading ones can be treated along the same lines. Remember that in our conventions the $z$ dependence flows along the shortest path between the two shifted legs.

\begin{figure}[h!]
\centering
 \includegraphics[scale=0.5]{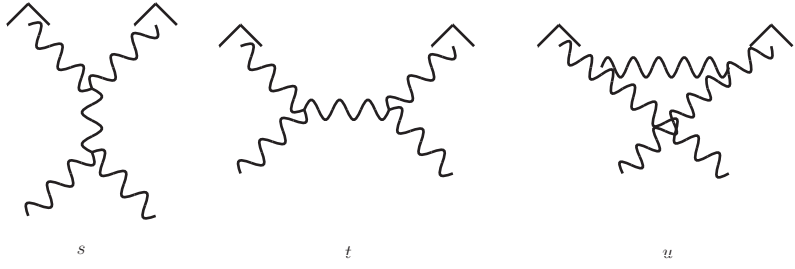}
 \caption{$s$, $t$, and $u$ channel cubic graphs used in the four point example. Hats denote the BCFW-shifted legs.}
\end{figure}

\subsubsection*{Example: four point Feynman graphs}
At four points tree level there are only three cubic diagrams to consider: $s$, $t$, and $u$-channel with the four-vertex absorbed into the cubic graphs according to the color factors.  In the BCJ representation this is written as
\begin{equation}
\mathcal{A}_4= \frac{c_s n_s}{s}+\frac{c_t n_t}{t}+\frac{c_u n_u}{u}
\end{equation}
In the following two particle momenta will be BCFW shifted and the other two legs will remain off-shell. In this way the analysis extends to the integrand. More precisely it extends to the Feynman graphs of the integrand with maximally one hard propagator in between the shifted legs. For graphs with more hard propagators the power counting will become more complicated as the number of graphs increases but in spirit it is similar. Hence, on the level of the integrand the leading large-$z$ behavior for this class of diagrams is in a BCJ form given by
\begin{equation}\label{tri4points}
 \mathcal{I}_4 = \frac{c_s n_s}{s}+\frac{c_t n_t}{t}+\frac{c_u n_u}{u}
\end{equation}
with
\begin{equation}
 c_u=c_s-c_t \quad \Rightarrow \quad n_u=n_s-n_t
\end{equation}
Writing down the Feynman graphs and shifting particles $1$ and $2$ one can extract the large-$z$ behavior of the kinematic numerators straightforwardly and it follows for the leading power in $z$ for the three numerators (with indices $\sigma$ and $\rho$ meaning to be contracted into appropriate currents)
\begin{equation}
 \begin{gathered}
   \lim_{z\rightarrow \infty} n_s=\varepsilon(\hat{1})_\mu \varepsilon(\hat{2})_\nu
\Big(4iz\eta^{\mu\nu}\eta^{\rho\sigma}p_4\cdot q+\mathcal{O}(z^0)\Big)\\
   \lim_{z\rightarrow \infty} n_t=\varepsilon(\hat{1})_\mu \varepsilon(\hat{2})_\nu
\Big(2iz(\eta^{\mu\sigma}\eta^{\nu\rho}-\eta^{\mu\rho}\eta^{\nu\sigma}+\eta^{\mu\nu}\eta^{
\rho\sigma}) p_4\cdot q +\mathcal{O}(z^0)\Big)\\
   \lim_{z\rightarrow \infty} n_u=\varepsilon(\hat{1})_\mu \varepsilon(\hat{2})_\nu
\Big(2iz(-\eta^{\mu\sigma}\eta^{\nu\rho}+\eta^{\mu\rho}\eta^{\nu\sigma}+\eta^{\mu\nu}\eta^{
\rho\sigma
} )p_4\cdot q +\mathcal{O}(z^0)\Big)
 \end{gathered}
\end{equation}
While these numerators satisfy the Jacobi relation above at leading order in $z$ (this is special for four points) they do not yet have the form one would expect from the discussion in the previous sections, i.e $z$ times a metric between the shifted legs. To get to this form one has to make use of the color Jacobi relations. It can be used to shift all terms not proportional to the metric from $n_u$ to $n_t$ and $n_s$ in \eqref{tri4points}. In the former numerator these will cancel with the other terms not proportional to $\eta^{\mu\nu}$ and in the latter numerator these terms will become subleading in the large-$z$ limit. This corresponds to shifting terms from the four-vertex between different kinematic channels. After some algebraic gymnastics one arrives at
\begin{equation}
  \begin{gathered}
   \lim_{z\rightarrow \infty} n_s=\varepsilon(\hat{1})_\mu \varepsilon(\hat{2})_\nu
\Big(4iz\eta^{\mu\nu}\eta^{\rho\sigma}p_4\cdot q+\mathcal{O}(z^0)\Big)\\
   \lim_{z\rightarrow \infty} n_t=\varepsilon(\hat{1})_\mu \varepsilon(\hat{2})_\nu
\Big(2iz\eta^{\mu\nu}\eta^{
\rho\sigma} p_4\cdot q +\mathcal{O}(z^0)\Big)\\
   \lim_{z\rightarrow \infty} n_u=\varepsilon(\hat{1})_\mu \varepsilon(\hat{2})_\nu
\Big(2iz\eta^{\mu\nu}\eta^{
\rho\sigma
} p_4\cdot q +\mathcal{O}(z^0)\Big)
 \end{gathered}
\end{equation}
which nicely mirrors (leading part) of the result \eqref{numtreelevel} obtained using pseudoinverses. It can be shown in a similar way that the subleading pieces are antisymmetric.
\subsubsection*{Higher points and inclusion of matter}
In principle this procedure can be done in a similar fashion at higher points including the subleading parts. As the number of Jacobi relations increases quite rapidly with the number of points this becomes, however, more and more intricate. We could reproduce the scaling \eqref{numtreelevel} in this direct Feynman graph approach up to including six points and our suspicion is that this can be done at any number of points. Furthermore, similar steps can be repeated for scalars and fermions in the adjoint and one arrives at the same result as before. This was also checked up to including six points.

\subsection{BCFW shifts versus color-kinematic duality}
The problem with the direct approach using Feynman diagrams is that the numerators obtained in this way do not satisfy the Jacobi relations beyond four points directly, i.e. they obey
\begin{equation}
 \{n_i-n_j+n_k=u_a,\; a=1,...,\# Jacobis\}
\end{equation}
where the right-hand side is non-vanishing and different for each Jacobi relation. One can now make use of gauge transformations on the numerators to bring them into a BCJ satisfying form, implementing a numerator shift
\begin{equation}
 n_i\rightarrow n_i+\Delta_i \quad\text{with}\quad \sum_i\frac{c_i\Delta_i}{s_i}=0 \;\;\; \forall i
\end{equation}
For instance at tree level this works at follows: Following \cite{Bern:2008qj} there are $(2n-5)!!$ kinematic numerators / color factors at $n$ points tree level. These obey $\frac{(n-3)(2n-5)!}{3}$ Jacobi relations which involve quite a lot of redundancy so that the number of non-redundant Jacobi relations is given by $(2n-5)!!-(n-2)!$. In other words $(n-2)!$ kinematic numerators / color factors are independent. Having constructed the numerators using Feynman diagrams one would have consequently obtained $(2n-5)!!-(n-2)!$ non-redundant Jacobi relations that are not satisfied. 
\begin{equation}
 \{n_i-n_j+n_k=u_a,\; a=1,...,(2n-5)!!-(n-2)!\}
\end{equation}
By enforcing a generalized gauge transformation on all $(2n-5)!!$ kinematic numerators the right hand side of the non-redundant Jacobi relations can be brought to zero by requiring (in addition to the gauge condition above)
\begin{equation}
 \begin{split}
 \{\Delta_i-\Delta_j+\Delta_k=u_a, a=1,...,(2n-5)!!-(n-2)!\}
 \end{split}
\end{equation}
so that the non-vanishing right side of the kinematic Jacobi relations gets cancelled. Based on pure counting the generalized gauge condition would give $(n-2)!$ conditions on the shifts so that one the number of total equations for the shifts $\Delta$ is
\begin{equation}
 \underbrace{(2n-5)!!-(n-2)!}_{\# \text{Jacobi relations to bring to zero}} +
\underbrace{(n-2)!}_{\#\text{ gauge conditions}} = (2n-5)!! 
\end{equation}
i.e. $(2n-5)!!$ equations for $(2n-5)!!$ shifts which means that in principle there should be a solution to this system of equation. This of course is hard to handle already at five points as the expressions involved become quite unhandy. Moreover, this raises the question if applying the generalized gauge transformations that bring the numerators into a BCJ satisfying form respects the large-$z$ scaling? This question cannot be satisfyingly answered for arbitrary multiplicity using this approach but we have checked that (at least) up to five points shifts can be found and are such that they do not spoil the large-$z$ behavior but in light of the discussion in the previous sections it is quite likely that such shifts exist for higher multiplicity as well. Of course, the same arguments go through at the level of the integrand.


\section{Discussion and conclusion}\label{sec:discussion}
The interaction of color-kinematic duality and powercounting has been studied in gauge and gravity theories. The general analysis has been applied in two related but logically distinct contexts: that of the large BCFW shift and that of large loop momenta. The main message of this article is that (up to assumptions that were addressed) a reformulation of color-kinematics duality in terms of linear maps is neatly suited to refine powercounting in gravity theories thereby making cancellations manifest. In this way, we have first derived the large-BCFW scaling of quite general gravity theory integrands and could show that it scales drastically better than  na\"ively powercounting Feynman graphs leads to suspect. Secondly, we used the linear map approach to UV divergences of gravity theories and could interpret the results of Zvi Bern \emph{et al.} for $\mathcal{N}=8$ supergravity computations from a powercounting perspective. Moreover, our analysis seems to indicate that for $\mathcal{N}=8$ to have the same powercounting behavior as $\mathcal{N}=4$ SYM one should include tadpole graphs at $5$ loops in $\mathcal{N}=4$. In general, the reformulation of powercounting through color-kinematics and linear maps opens a window to a far more general approach than hitherto employed.

The problem of proving color-kinematic duality remains open at loop level. Proving this is crucial to strengthen the validity of the results of this article. The linear map approach explored above might give a natural starting point for this. This analysis of color-kinematic duality should also provide a clear answer what role the vanishing terms in the integrand play. Moreover, it should include a discussion of which trivalent graphs to include in the construction. Both of these play an important role in the analysis of the UV divergences presented above.  This obviously is also a direction for future research to decide the question if $\mathcal{N}=8$ is a finite quantum field theory. 

There are many interesting questions to guide further research which lead from this work. For instance, it is easy to suspect but hard to make precise that the fall-off for gravity integrands obtained above implies the existence of on-shell recursion relations for the gravity integrand. At the very least, it implies the gravity integrand can be reconstructed from its single cut singularities, but unfortunately these have never been identified in terms of lower loop integrands except in the special case of $\mathcal{N}=4$ super Yang-Mills theory \cite{ArkaniHamed:2010kv}. As an extension of this, one can ask the question if the $\frac{1}{z^2}$ fall-off for gravity amplitudes imply the existence of bonus relations for gravity integrands, along the same lines as \cite{Spradlin:2008bu} at tree level. Next, one could wonder if counterterms in theories which do have UV divergences have a non-trivial interaction with color-kinematic duality as suggested by the results of \cite{Broedel:2012rc}.

The techniques explored in this article have the potential to sharpen the analysis of perturbative quantum gravity considerably. As color-kinematics duality offers a promising path towards understanding quantum gravity, further study would be most welcome and very interesting.

\section*{Acknowledgements}
It is a pleasure to thank Zvi Bern and John-Joseph Carrasco for correspondence and discussions respectively. In addition, we would like to thank Donal O'Connell, Ricardo Monteiro and Gang Yang for discussions and collaboration on related projects. Moreover, we would like to thank the anonymous referee for helpful suggestions. RB is grateful to the Institute for Advanced Study for hospitality while this article was begun being finished. This work has been supported by the German Science Foundation (DFG) within the Collaborative Research Center 676 ``Particles, Strings and the Early Universe".  Jaxodraw \cite{Binosi:2008ig} based on Axodraw \cite{Vermaseren:1994je} has been used to produce the figures.

\pagebreak
\begin{appendices}
\section{Feynman rules in AHK gauge} \label{app:feynrulesAHK}
\begin{figure}[h!]
\centering
\includegraphics[scale=0.8]{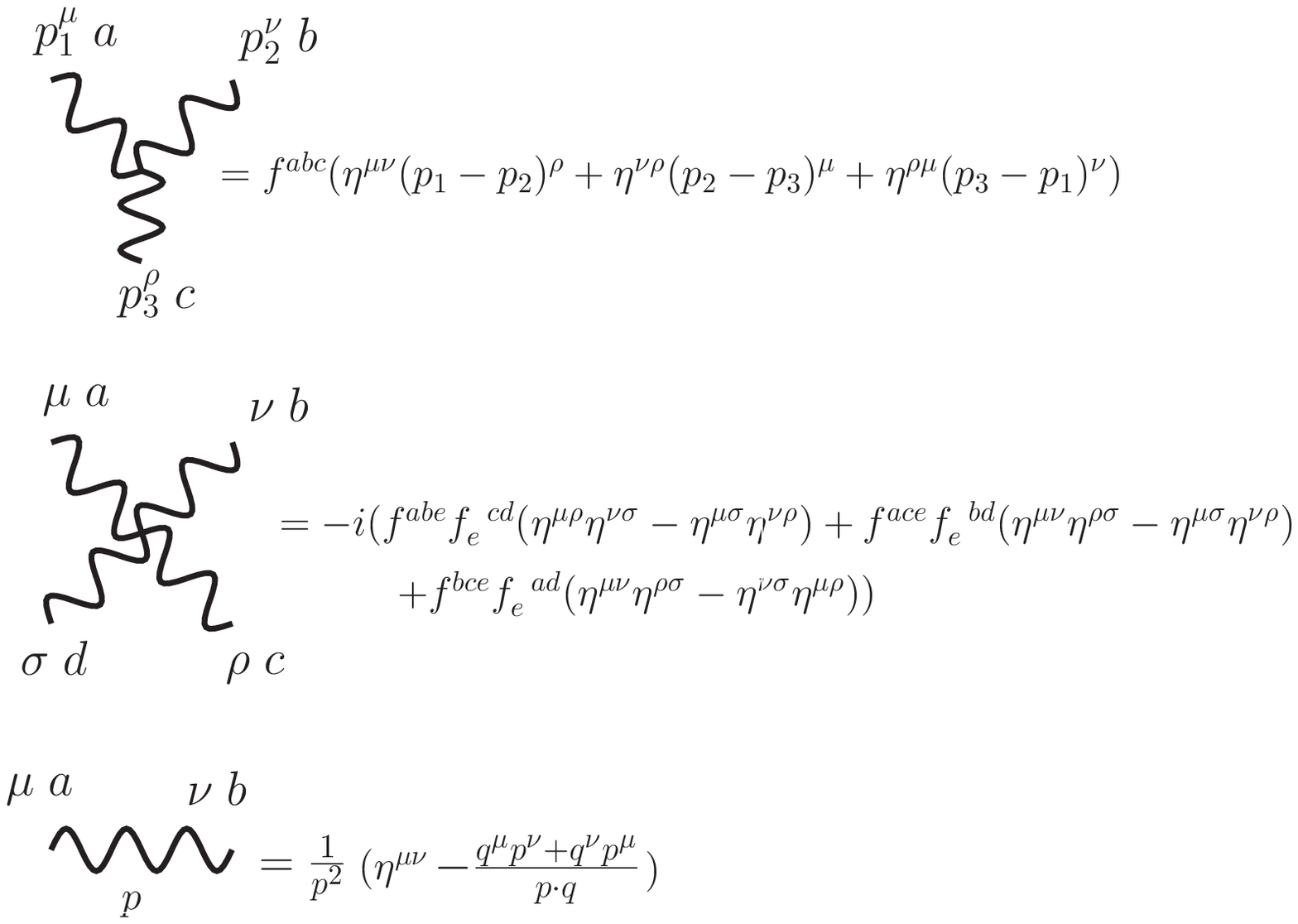} \label{fig:powercountingfeynmanrules}
 \caption{Color-ordered Feynman rules used for power counting in chapter \ref{sec:feynman}. The vertices are in Feynman-'t Hooft gauge and the propagator in AHK $A\cdot q=0$ lightcone gauge.}
\end{figure}

\section{The Moore-Penrose pseudoinverse}\label{sec:pseudoinv}
In this paper only the defining property  \eqref{consitencygeninv} of the generalized inverse is needed, as well as its ability to solve linear equations. In the literature more specific notions of generalized inverse abound which satisfy various other properties in addition to the defining equation \eqref{consitencygeninv}. A prominent example of this is the Moore-Penrose pseudoinverse  \cite{Moore1920,Penrose:1955vy}.  This is a generalized inverse of a matrix $A \in \mathbb{C}^{m\times n}$ denoted by $X$ which satisfies
\begin{enumerate}
 \item $AXA=A$ (The definition of the generalized inverse)
 \item $XAX=X$ 
 \item $(XA)^*=AX$ ($XA$ is hermitian)
 \item $(AX)^*=XA$ ($AX$ is hermitian)
\end{enumerate}
It can be shown \cite{Moore1920,Penrose:1955vy} that the solution to these conditions is unique. Other notions of generalized inverse impose a subset of these conditions. In practice the Moore-Penrose pseudoinverse can be computed using the singular value decomposition, i.e. one can factorize the matrix $A$ as
\begin{equation}\label{eq:SVDpsuedo}
 A=U\Sigma V^*
\end{equation}
where $U$ is a $m\times m$ complex unitary matrix, $\Sigma$ is an $m \times n$ rectangular diagonal matrix with nonnegative real numbers on the diagonal, and $V^*$  is the conjugate transpose of a matrix V, i.e. an $n\times n$ real or complex unitary matrix. Then the pseudoinverse is given by
\begin{equation}
 X=V\Sigma^+ U^*
\end{equation}
with $\Sigma^{-1}$  the pseudoinverse of $\sigma$, formed by replacing every nonzero diagonal entry by its reciprocal and then taking its transpose. Conveniently, Mathematica has the built-in functions \textit{SingularValueDecomposition} and \textit{PseudoInverse} to calculate Moore-Penrose inverses. 

\section{Solving Jacobi relations: minimizing distances} \label{app:mindist}
In section \ref{sec:rev} it was shown how the color Jacobi relations could be solved while maximizing the distance between two fixed labels. A natural second solution follows by using the Jacobi relation to \emph{minimize} the number of structure constants encountered from $1$ to $n$. To see how this works consider for instance the structure constants in \eqref{eq:minbasisI}. Now the Jacobi relation can be used to relate all different orderings in this set to one another, at the cost of introducing structure constants with less distance between $1$ and $n$. It will be advantageous to aim for the natural minimal lexicographic ordering although this is not completely possible beyond four points. As a first step consider particle $2$ and an arbitrary but fixed ordering of the remaining $n-3$ particles. The Jacobi relation can be used to shuffle particle 2 to the minimal lexicographically ordered position, i.e.
\begin{equation}
F(1,a_1,a_2,\ldots,a_i,2, \ldots, n) = F(1,2,a_1,a_2,\ldots, n) - \sum_{j=1}^i \sum_e F(1,a_1,\ldots,e,\ldots,n) F(e,a_j,2)
\end{equation}
at the cost of introducing lesser length structure constant contractions. Now the would-be basis consists of one structure constant contraction of maximal length $n-3$ and structure constants of lesser length. The same procedure can now be repeated at these lower length orders as well. There are now two complications. First, as mentioned before it is not possible to express these lesser length structure constants in terms of lexicographically ordered sets only. To see this consider for instance five points: there are in total six independent color structures but only five minimally lexicographically ordered ones. Secondly, there are now external sub-graphs attaching to the main line. The number of these are determined by applying Jacobi-relations in these subgraphs as well: for our purposes only the distance between the particles $1$ and $n$ will be important. 

Explicitly, consider structure constants contractions with a distance $k$ between the special particles $1$ and $n$ on an $n$ point amplitude. First consider the case where there are $k$ legs with one and 1 leg with $n-k-2$ legs. For the latter there are $(n-k-3)!$ ways to attach structure constants modulo the Jacobi identity. Then one can consider the case with $k-2$ legs with one and two legs with $n-k$ particles in total. All different possibilities count. 

As a check that this second basis is minimal one can calculate its dimension. Counting the number of basis elements of a distance of $k$ between the special particles $1$ and $n$ for an n-point amplitudes in the second basis can be mapped to the problem of counting the number of possibilities of having a permutation of $(n-2)$ elements with $k$ different cycles. This number is the unsigned stirling number of the first kind, $s(n-2,k)$. For these numbers
\begin{equation}
\sum_{k=1}^{n-2} s(n-2,k) = (n-2)!
\end{equation}
holds. Hence the second bases has the same dimension as the first one. 

Using either basis one can express the amplitude as a sum over basis elements $f^i$. Then the right and left hand equation \eqref{eq:BCJtree} as a sum over these
\begin{equation}\label{eq:a=b}
\sum_{i}  \alpha_i f^i = \sum_i \beta_i f^i
\end{equation}
Since the basis elements are independent, this constitutes $(n-2)!$ equations in $(2n - 5)!!$ unknown numerators \cite{Kleiss:1988ne} which is always solvable. This shows explicitly there is a large degree of freedom in choosing the numerators. This is the freedom which is probed by the generalized gauge transformation. 

To put differently, the coefficients $\beta_i$ on the right hand side of equation \eqref{eq:a=b} are $n-2$ functions of the numerators. For any set of numerators for which $\beta_i(n_j) = \alpha_i $, equation \eqref{eq:BCJintegrand} gives the tree level amplitude $A$. 

\end{appendices}



\bibliographystyle{jhep}

\bibliography{gravityshiftsbib}

\end{document}